\documentclass[nofootinbib]{ws-book9x6}
\usepackage{ws-book-thm}   
\usepackage{ws-book-har}   
\usepackage[pdfpagelabels=false]{hyperref}  
\usepackage{amsfonts}
\usepackage{amsmath}
\usepackage{amssymb}
\usepackage{bm}
\usepackage{dcolumn}
\usepackage{graphicx}
\usepackage{graphics}
\usepackage{latexsym}
\usepackage{rotating}
\usepackage[all]{hypcap} 
\usepackage{xspace} 
\usepackage[usenames]{xcolor}
\usepackage{mathrsfs}
\usepackage{multirow}

\title{Topics on Strong Gravity}      

\makeindex

\def\apj{Astrophys. J. \ }%
\def\aap{A\&A}%
\begin{document}

\setcounter{page}{1}

\chapter[Astrophysical aspects of general relativistic mass twin stars]
        {Astrophysical aspects of general relativistic mass twin stars }
\label{ch1}

\begin{centering}
David Blaschke$^{1,2,3}$, 
David Edwin Alvarez-Castillo$^{1}$,  \\
Alexander Ayriyan$^{4,5}$, 
Hovik Grigorian$^{4,5,6}$,
Noshad Khosravi Lagarni$^{1,7}$,\\
Fridolin Weber$^{8,9}$
\\[5mm]

{\it
$^{1}$ Bogoliubov Laboratory for Theoretical Physics,
	Joint Institute for Nuclear Research,
	Joliot-Curie street 6,
	141980 Dubna, Russia\\
$^{2}$ Institute of Theoretical Physics, 
	University of Wroclaw, 
	Max Born place 9, 
	50-204 Wroclaw, Poland\\
$^{3}$ National Research Nuclear University (MEPhI),
	Kashirskoe Shosse 31,
	115409 Moscow, Russia\\
$^{4}$ Laboratory for Information Technologies,
	Joint Institute for Nuclear Research,
	Joliot-Curie street 6,
	141980 Dubna, Russia\\
$^{5}$ Computational Physics and IT Division, 
	A.I. Alikhanyan National Science Laboratory, 
	Alikhanyan Brothers street 2, 
	0036 Yerevan, Armenia\\	
$^{6}$ Department of Physics, 
	Yerevan State University, 
 	Alek Manukyan street 1, 
 	0025 Yerevan, Armenia\\
$^{7}$ Department of Physics, 
        Alzahra University, 
        Tehran, 1993893973, Iran\\
$^{8}$ Department of Physics, 
        San Diego State University, 
        5500 Campanile Drive, 
        San Diego, CA 92182, USA\\ 
$^{9}$ Center for Astrophysics and Space Sciences,
        University of California,
        San Diego, La Jolla, CA 92093, USA
}

\vspace{1cm}

\begin{minipage}{10cm}
Abstract. In this chapter we will introduce an effective equation of state (EoS) model based on polytropes that serves to
study the so called "mass twins" scenario, where two compact stars have approximately the same mass but (significant for observation) quite different radii. Stellar mass twin configurations are obtained if a strong first-order phase transition occurs in the interior of a compact star. In the mass-radius diagram of compact stars, this will lead to a third branch of gravitationally 
stable stars with features that are very distinctive from those of white dwarfs and neutron stars.
We discuss rotating hybrid star sequences in the slow rotation approximation and in full general relativity and draw conclusions for an upper limit on the maximum mass of nonrotating compact stars that has recently be deduced from the observation of the merger event GW170817.
\end{minipage}

\end{centering}

\vspace{1cm}

\section{Introduction}
\label{intro}

Compact stars, the stellar remnants following the death of main
sequence stars, have been the subject of investigation since the
beginning of the last century. In particular, the determination of the
internal composition of neutron stars is an open problem. Researching
it involves many areas of physics, like nuclear, plasma, particle
physics and relativistic astrophysics. Moreover, due to the enormous
compactness (as expressed in the mass-radius ratio) of compact stars,
these objects are extremely relativistic. Therefore, one can neither
exclusively apply non-relativistic quantum mechanics nor classical
Newtonian gravity to describe the observational properties of compact
stars.

During the last decade important astronomical observations have shed light onto the
nature of the dense, cold matter in the stellar interiors of compact
stars. The detection of massive neutron stars, of about $2 M_{\odot}$,
has constrained the maximum density values in their cores and also
revealed the stiff nature of the nuclear equation of state (EoS) at
ultra-high densities. Strongly related to this issue, and one of the
most interesting aspects of modern dense-matter physics, concerns the
possible onset of quark deconfinement in the cores of compact stars.


Microscopic models that take into account the nuclear interactions
either at the nucleon or quark level aim at providing a realistic
hadronic or quark matter EoS, respectively.  Neutron star matter must
be thermodynamically consistent.  Interestingly, due to the fast
cooling of neutron stars after their birth in a supernova collapse the
thermal contributions to the EoS do not contribute substantially and
can safely be neglected \cite{Yakovlev:2000jp}.  The thermodynamic
system can therefore be described by three macroscopic variables:
energy density $\varepsilon$, baryonic density $n$, and pressure $P$.
A fourth quantity of great interest, the chemical potential, can then
be obtained as $\mu=(P+\varepsilon)/n$.

In addition, the most basic conditions that the system must fulfill
include global charge neutrality and $\beta$-equilibrium.  The latter
is derived from the reaction balance of beta decay and its inverse,
the electron capture, due to the weak interactions and fixes the
relation between the chemical potentials of different species in the
system.

With the above conditions satisfied, the neutron star equation of
state becomes an expression of the form $P(\varepsilon)$, where
$\varepsilon=\varepsilon(n)$ and $P=P(n)$ acquire parametric
forms. Furthermore, in order to compute the internal properties of
compact stars, it is necessary to obtain internal pressure profiles.
This will result in mass-radius relations that characterize an EoS.
Neutron stars are extremely relativistic objects which require to be
treated within Einstein's general theory of relativity rather than
simply Newtonian gravity, which may still be applicable to white dwarf
stars.  In this sense, our contribution addresses "strong gravity" in
a unique fashion.  To give an example, for a pulsar of mass
$M=2~M_\odot$ with a typical radius of around 12~km, 
{the general relativistic correction factor amounts to $1/(1-2GM/R) = 2$
\cite{Tolman:1939jz,Oppenheimer:1939ne}, which is a 100\% correction
relative to Newtonian gravity!}

In the following sections we will introduce an effective EoS model
based on polytropes~\citep{Alvarez-Castillo:2017qki} that serves to
{study} the so called "mass twins" scenario, where two
compact stars have approximately the same mass 
{but  (significant for observation) quite different radii
\cite{Glendenning:1998ag}.}  
Stellar mass twin configurations are obtained if a strong first-order phase transition
occurs in the interior of a compact star. In the mass-radius diagram
of compact stars, this will lead to a third branch of gravitationally
stable stars with features that are very distinctive from those of
white dwarfs and neutron stars. 
The condition on the EoS that will lead to mass twins was first derived by 
Seidov~\citep{1971SvA....15..347S}, 
see also \citep{1983A&A...126..121S,Lindblom:1998dp},
namely that the central energy
density $\varepsilon_v$, central pressure $P_c$, and the jump in
energy associated with the phase transition $\Delta\varepsilon$ obey the
relation
\begin{eqnarray}
\label{seidov}
\frac{\Delta\varepsilon}{\varepsilon_c}\ge\frac{1}{2} +
\frac{3}{2}\frac{P_c}{\varepsilon_c} \, .
\end{eqnarray}
When fulfilled, the corresponding compact star will suffer an
instability of the same type as the maximum mass star of a stellar
sequence. Most interestingly, stars with central densities higher than
the density of the maximum-mass star become stable again if their
gravitational masses obey $\partial M/\partial\varepsilon_c(0)>0$,
thereby populating a third branch with stable mass twins.
 
\section{Self-consistent set of field equations for stationary rotating and
tidally deformed stars}
\label{gravity}

The geometrical description of the space-time structure
curved due to the mass -energy of a compact star is given by the general
metric form defining the interval between the infinitesimally close
events,
\begin{eqnarray}
ds^{2} & = & g_{\mu\nu}(x)dx^{\mu}dx^{\nu} ~ .
\label{metric}
\end{eqnarray}
The curvature of the space-time is satisfying the Einstein field equations
$G_{\nu}^{\mu}=8\pi GT_{\nu}^{\mu}, $where $G_{\nu}^{\mu}=R_{\nu}^{\mu}-1/2\delta_{\nu}^{\mu}R$.
Here $R_{\nu}^{\mu}$ is the Ricci curvature tensor and $R$ the scalar
curvature. 
On the right hand side of the Einstein equation we have
the energy -momentum tensor $T_{\nu}^{\mu}$ of the stellar matter
and $G$ is the gravitational constant ($\hbar=c=1)$.

The metric tensor $g_{\mu\nu}(x)$ has the same symmetry as the matter
distribution. Therefore, if one assumes that the star is static and
not deformed the metric tensor is diagonal and depends only on the
distance from the center of the star.  In the case of stationary
rotating stars the symmetry of the matter will be axial symmetric.  In
this case due to the rotational motion of the star the non-diagonal
element $g_{t\phi}$ ($t$-time coordinate, $\phi$ - azimutal angle of
the spherical coordinate system) will not be zero in the inertial
frames connected with the star.  The existence of such a term leads to
the Lense-Thirring effect of frame dragging for the motion of bodies
in the gravitational field of rotating compact relativistic stellar
objects.  However, in the case of small but static deformations the
metric will be non-spherical but diagonal.

\subsection{Einstein equations for axial symmetry}
\label{gravity:Einstein}

The general form of the metric for an axially symmetric space-time manifold
in the inertial frame where the star center is at rest is 
\begin{eqnarray}
\label{metric1}
ds^{2} & = &
e^{\nu(r,\theta)}dt^{2}-e^{\lambda(r,\theta)}dr^{2}-r^{2}e^{\mu(r,\theta)}[d\theta^{2}+\sin^{2}
  \theta(d\phi+\omega(r,\theta)dt)^{2}], \nonumber\\
\label{metric-1}
\end{eqnarray}
where a spherically symmetric coordinate system has been used in order to obtain
the Schwarzschild solution as a limiting case. 
This line element is time - translation and axial-rotational invariant; all metric functions
are dependent on the coordinate distance from the coordinate center
$r$ and altitude angle $\theta$ between the radius vector and the
axis of symmetry.

The energy momentum tensor of stellar matter can be approximated by
the expression of the energy momentum tensor of an ideal fluid 
\begin{equation}
T_{\mu}^{\nu}=(\varepsilon+P)u_{\mu}u^{\nu}-P\delta_{\mu}^{\nu},\label{Energy}
\end{equation}
where $u^{\mu}$ is the $4$-velocity of matter, $P$ the pressure
and $\varepsilon$ the energy density.

Once the energy-momentum tensor (\ref{Energy}) is fixed by the choice
of the equation of state for stellar matter, the unknown metric functions
$\nu$,$\lambda$, $\mu$, $\bar{\omega}$ can be determined by the
set of Einstein field equations for which we use the following four
combinations.

There are three Einstein equations for the determination of the diagonal
elements of the metric tensor, 
\begin{eqnarray}
\label{einstein}
G_{r}^{r}-G_{t}^{t} & = & 8\pi G(T_{r}^{r}-T_{t}^{t})~,
\\
G_{\theta}^{\theta}+G_{\phi}^{\phi} & = & 8\pi G(T_{\theta}^{\theta}+T_{\phi}^{\phi})~,
\\
G_{\theta}^{r} & = & 0~,
\end{eqnarray}
and one for the determination of the non-diagonal element 
\begin{equation}
\label{rote}
G_{\phi}^{t}=8\pi GT_{\phi}^{t}~.
\end{equation}
We use also one equation for the hydrodynamic equilibrium (Euler equation)
\begin{equation}
{H}(r,\theta)\equiv\int\frac{dP^{\prime}}{P^{\prime}+\varepsilon^{\prime}}
=\frac{1}{2}\ln[u^{t}(r,\theta)]+{\rm const},\label{euler}
\end{equation}
where the gravitational enthalpy ${H}$ thus introduced is a function
of the energy and/or pressure distribution.

\subsection{Full solution for uniform rotational bodies}
\label{gravity:RNS}

In this section we describe the method of solution employed by the RNS code written by~\citet{Stergioulas:1994ea}, 
based on the method developed by~\citet{Komatsu:1989zz} that also includes modifications by 
\citet{Cook:1993qj}. 
In addition, the inclusion of quadrupole moments is due to Morsink based on the method by~\citet{Laarakkers:1997hb}.

In order to study the full solutions for rotating compact stars the following metric is considered~\cite{Cook:1993qj}: 
\begin{eqnarray}
ds^{2} & = &
-e^{\gamma(r,\theta)+\rho(r,\theta)}dt^{2}+e^{2\alpha(r,\theta)}(dr^{2}+r^{2}d\theta^{2})
+e^{\gamma(r,\theta)-\rho(r,\theta)}r^{2}\sin^{2}\theta  \nonumber \\
&&\times ~(d\phi-\omega(r,\theta)
dt)^{2},
\end{eqnarray}
which just like Eq.~(\ref{metric1}) properly describes a stationary,
axisymmetric spacetime.  In addition, the matter source is chosen to
be a perfect fluid described by the stress energy tensor
$T^{\mu\nu}=(\epsilon+P)u^{\mu}u^{\nu}+Pg^{\mu\nu}$, where $u^{\mu}$
is the four-velocity of matter.  Three of the solutions to the
gravitational field equations are found by a Green function approach
therefore leading to the determination of the metric potentials
$\rho$, $\gamma$ and $\omega$ in term of integrals, whereas the
$\alpha$ potential is found by solving a linear differential equation.
Therefore, we find the corresponding numerical solutions to our compact star
models by employing the RNS code. 
In the formulation of the problem, all the physical variables are written in dimensionless form by means
of a fundamental length scale $\sqrt{\kappa}$, where $\kappa \equiv
\frac{c^{2}}{G\epsilon_{0}}$ with $\epsilon_{0}\equiv
10^{15}$g~cm$^{-3}$.
\begin{table}[!htb]
\tbl{Dimensionless physical parameters of the gravitational field
  equations used in the RNS code formulation.}{ \centering
\begin{tabular}{|c|c|c|c|c|c|c|c|c|}
\hline 
\hline 
$\bar{r}$  & $\bar{t}$  & $\bar{\omega}$  & $\bar{\Omega}$  & $\bar{\rho_{0}}$  & $\bar{\epsilon}$  & $\bar{P}$ & $\bar{J}$ &$\bar{M}$ \\
\hline
$\kappa^{-1/2}r$  & $\kappa^{-1/2}ct$  & $\kappa^{1/2}\frac{1}{c}\omega$  & $\kappa^{1/2}\frac{1}{c}\Omega$  & $\kappa\frac{G}{c^{2}}\rho_{0}$  & $\kappa\frac{G}{c^{2}}\epsilon$  & $\kappa\frac{G}{c^{4}}P$ & $\kappa^{-1}\frac{G}{c^{3}}J$ &$\kappa^{-1/2}\frac{G}{c^{2}}M$ \\
 \hline 
\end{tabular}
}
\label{scaled_param}  
\end{table}
\begin{table}[!htb]
\tbl{Output parameters of the RNS code.} {%
\centering
\begin{tabular}{|l|l|}
\hline 
\hline 
Gravitational mass-energy  & $M/M_{{\odot}}$  \tabularnewline
Rest mass & $M_{0}/M_{\odot}$ \tabularnewline
Circunferencial radius [km] & $R_{e}$ \tabularnewline
Eccentricity& $e$ \tabularnewline
Central energy density [10$^{15}$ g~cm$^{-3}$] & $\epsilon_{c}$ \tabularnewline
Angular velocity measured at infinity [10$^{3}$ s$^{-1}$] & $\Omega$ \tabularnewline
Total angular momentum & $cJ/GM^{2}_{\odot}$ \tabularnewline
Rotational kinetic energy over gravitational energy & $T/W$ \tabularnewline
Measure of frame dragging & $\omega_{c}/\Omega_{c}$ \tabularnewline
Polar redshift & $Z_{p}$ \tabularnewline
Equatorial redshift in backward direction & $Z_{b}$ \tabularnewline
Equatorial redshift in forward direction & $Z_{f}$ \tabularnewline
Circumferential height of corotating marginally stable orbit [km]& $h_{+}$ \tabularnewline
Circumferential height of counterrotating marginally stable orbit [km] & $h_{+}$ \\
\hline 
\end{tabular}} 
\label{RNS_output}  
\end{table}

The global parameters of the star are computed by means of the following expressions:

\begin{eqnarray}
  M &=& \frac{4\pi\kappa^{1/2}c^{2}\bar{r}^{3}_{e}}{G}\int^{1}_{0}
  \frac{s^{2}ds}{(1-s)^{4}} \int^{1}_{0}d\mu~e^{2\alpha+\gamma}
  \nonumber \\ && \times ~\left\{ \frac{\bar{\epsilon}+\bar{P}}{1-v^{2}} \left[
    1+v^{2}+\frac{2sv}{1-s}(1-\mu)^{1/2} \hat{\omega}e^{-\rho}
    \right]+ 2\bar{P} \right\}, \\
%
  M_{0}&=&\frac{4\pi\kappa^{1/2}c^{2}\bar{r}^{3}_{e}}{G}\int^{1}_{0}\frac{s^{2}ds}{(1-s)^{4}}
  \int^{1}_{0}d\mu~e^{2\alpha+(\gamma-\rho)/2}
  \frac{\bar{\rho_{0}}}{(1-v^{2})^{1/2}}, \\ 
%
  J  &=& \frac{4\pi\kappa
    c^{3}\bar{r}^{4}_{e}}{G}\int^{1}_{0}\frac{s^{3}ds}{(1-s)^{5}}
  \int^{1}_{0}d\mu (1-\mu^2)^{1/2}~e^{2\alpha+\gamma-\rho}
  (\bar{\epsilon}+\bar{P}) \frac{v}{1-v^{2}}, \nonumber \\ & & \\
%
  T&=&\frac{2\pi\kappa^{1/2}c^{2}\bar{r}^{3}_{e}}{G}\int^{1}_{0}\frac{s^{3}ds}{(1-s)^{5}}
  \int^{1}_{0}d\mu (1-\mu^2)^{1/2}~e^{2\alpha+\gamma-\rho}
  (\bar{\epsilon}+\bar{P}) \frac{v\hat{\Omega}}{1-v^{2}}, \nonumber \\ & & 
\end{eqnarray}
where $\hat{\omega}\equiv\bar{r}_{e}\bar{\omega}$ and
$\hat{\Omega}\equiv\bar{r}_{e}\bar{\Omega}$ with $ \bar{r}_{e}$ as the
coordinate radius of the equator. Moreover, all the resulting
quantities can be written in terms of the auxiliary variables $\mu$
and s, defined as $\mu \equiv \theta$ and $\bar{r} \equiv
\bar{r}_{e}\left( \frac{s}{1-s} \right)$, respectively. Consequently,
the four metric functions acquire a dependence on the above variables:
$\rho(s,\mu)$, $\gamma(s,\mu)$, $\omega(s,\mu)$, $\alpha(s,\mu)$. The
remaining quantities are:
\begin{equation}
R_{e}=\kappa^{1/2}\bar{r}_{e}e^{(\gamma_{e}-\rho_{e})/2},
\end{equation}
\begin{equation}
Z_{p}=e^{-(\gamma_{p}+\rho_{p})/2}-1,
\end{equation}
\begin{equation}
Z_{f}=\left(\frac{1-v_{e}}{1+v_{e}}\right)^{1/2}~\frac{e^{-(\gamma_{e}+\rho_{e})/2}}{1+\hat{\omega}_{e}e^{-\rho_{e}}}-1,
\end{equation}
\begin{equation}
Z_{b}=\left(\frac{1+v_{e}}{1-v_{e}}\right)^{1/2}~\frac{e^{-(\gamma_{e}+\rho_{e})/2}}{1-\hat{\omega}_{e}e^{-\rho_{e}}}-1.
\end{equation}
where the subscripts $e$ and $p$ denote evaluation at the equation and at the pole, respectively.

For the solutions of maximally rotating compact stars in numerical general relativity the version of RNS code has been 
employed which was available for download by the time of the writing of this contribution from the website
\href{http://www.gravity.phys.uwm.edu/rns/}{http://www.gravity.phys.uwm.edu/rns/}

\subsection{Perturbation approach to the solution}
\label{gravity:mass}

The problem of the rotation can be solved iteratively by using a
perturbation expansion of the metric tensor and the physical
quantities in a Taylor series with respect to a small positive
parameter $\beta$. 
As such a parameter for the perturbation expansion
we use a dimensionless quantity. One of possible physically motivated
way is to take the ratio of the rotational or deformation energy to
the gravitational one. The gravitational energy could be estimated for
a homogeneous Newtonian star as $\beta=E_{\rm def}/E_{\rm grav}$. In
case of rotating stars the deformability is connected with the induced
centrifugal force and the expansion parameter is 
$\beta=E_{{\rm rot}}/E_{{\rm grav}}=(\Omega/\Omega_{0})^{2}$, where
$\Omega_{0}^{2}=4\pi G\rho(0)$ with the mass density $\rho(0)$ at the
center of the star. The choice of this parameter could be also
motivated with the conditions when the problem is discussed. For
example since for the stationary rotating stars can not have too high
value of the angular velocity, because of mass shedding on Keplerian
angular velocity $\Omega_{K}=\sqrt{GM/R_{e}^{3}}$ for the star with
total mass $M$ and $R_{e}$ equatorial radius, the expansion gives
sufficiently correct solutions already at $O(\Omega^{2})$. 
So the expansion parameter is naturally limited to values
$\Omega/{\Omega}_{0}\ll1$ by this condition of mechanical stability of
the rigid rotation, because always
$\Omega<\Omega_{K}={\Omega}_{0}/\sqrt{3}$. 
This condition is fulfilled
not only for homogeneous Newtonian spherical stars but also for the
relativistic configurations even with a possible hadron-quark
(deconfinement) transition, which we are going to discuss later in this
chapter, see Fig.~\ref{fig:MR_ACB4-ACB5_Pasta} below.
{
The perturbation approach to slowly rotating stars has been developed first 
by 
\citet{Hartle:1967he,1968ApJ...153..807H}, and 
independently by 
\citet{1968Ap......4...87S,1968Ap......4..227S}
and is described in detail in Refs.~\cite{Weber:1999Book,Glendenning:1997wn,Chubarian:1999yn}.
Our notation and derivation in this section will follow \citet{Chubarian:1999yn}
while numerical solutions for the slow rotation ($\Omega^2$-) approximation
are obtained with a code based on the improved Hartle scheme 
\cite{Weber:1992}.

The expansion of the metric tensor in a perturbation series with respect to the 
slow rotation parameter $\beta$ 
}
can be expressed as
\begin{equation}
g_{\mu\nu}(r,\theta)=\sum_{j=0}^{\infty}
\left(\sqrt{\beta}\right)^{j}g_{\mu\nu}^{(j)}(r,\theta)~.
\end{equation}
According to the metric form for the axial symmetry in the linear
approximation via $\beta$ parameter we introduce the notations
describing explicitly the non perturbed (spherically symmetric case
noted with upper index $(0)$) and perturbed terms (corresponding to
$j=1,2$) in the metric,
\begin{eqnarray}
\begin{array}{ccc}
\label{series}
e^{-\lambda(r,\theta)} & = & e^{-\lambda^{(0)}(r)}[1+\beta f(r,\theta)]+O(\beta^{2}) \, ,\\
e^{\nu(r,\theta)} & = & e^{\nu^{(0)}(r)}[1+\beta\Phi(r,\theta)]+O(\beta^{2}) \, , \\
e^{\mu(r,\theta)} & = & r^{2}[1+\beta U(r,\theta)]+O(\beta^{2}) \, ,
\end{array}
\end{eqnarray}
and for the frame dragging frequency $\omega$ the odd orders
\begin{eqnarray}
\begin{array}{ccc}
\label{intq}
\omega(r,\theta) & = & \sqrt{\beta}q(r,\theta)+O((\sqrt{\beta})^{3}) \, .
\end{array}
\end{eqnarray}

In the same way one performs a velocity expansion of the
energy-momentum tensor, the pressure and energy density distributions,
and of the kinetic energy,
\begin{eqnarray}
\label{euler1}
P(r,\theta) & = & P^{(0)}(r)+\beta
P^{(2)}(r,\theta)+O(\beta^{2})~,\\ \varepsilon(r,\theta) & = &
\varepsilon^{(0)}(r)+\beta\varepsilon^{(2)}(r,\theta)+O(\beta^{2}) \, ,
\end{eqnarray}
where $P^{(0)}$ and $\varepsilon^{(0)}$ denote the zero-order
coefficients which correspond to non-deformed spherically symmetric
stars.

Because of rotational symmetry the diagonal elements
$g_{\mu\mu}^{(j)}(r,\theta)$ (no summation over $\mu$) of the metric
coefficients can be written as ($j$ and $l$ are even values only)
\begin{eqnarray}
\label{Legendre-exp}
  g_{\mu\mu}^{(j)}(r,\theta) & = & \sum_{l=0}^{j}(g_{\mu\mu})_{l}(r)P_{l}(\cos\theta) \, .
\end{eqnarray}
The same is true also for the non diagonal elements, but the angular
dependence is different ($j$ and $l$ values are now odd only). The
case for the non diagonal term will be investigated in  section
\ref{ssec:moi}, where the  moment of inertia will be discussed.

In the case of a deformed distribution of the matter, the external metric
has the form of the Kerr metric. However this solution
is for black holes and does not correspond to a realistic stellar 
models. Therefore the external as well the internal solutions of the
field and matter distributions can only  be obtained either via a perturbation
approximation or via a completely numerical treatment.

\subsection{Static spherically symmetric star models}
\label{gravity:static}

{As a first step in the perturbation approach of a slowly
    rotating star one needs to find the internal gravitational field,
    the mass and matter distributions, the total gravitational mass,
    the radius and all other characteristic properties (including the
    metric functions) of spherically symmetric stars.}
The solution for the metric functions in  empty space (i.e., the
external solutions) are given by the Schwarzschild solution.

\begin{eqnarray}
\lambda^{(0)}(r) & = & -\ln[1-2GM/r],\label{phs-1}\\
\nu^{(0)}(r) & = & -\lambda^{(0)}(r).\nonumber 
\end{eqnarray}
where $M$ is a constant of integration, which asymptotically is
the Newtonian gravitational mass of the object.

These nonlinear equations, however, could be written in an elegant form
suggested by Tolman, Oppenheimer and Volkoff, which are known as the
TOV equations~\citep{Tolman:1939jz,Oppenheimer:1939ne}, and solved in
a way such that the internal solution matches the analytic external
Schwarzschild solution. The TOV equation is given by (for a
derivation, see, e.g., the textbook by 
\citet{Misner:1974qy})
\begin{equation}
\label{TOV}
\frac{dP^{(0)}(r)}{dr}=-G[P^{(0)}(r) +
  \varepsilon^{(0)}(r)]\frac{m(r)+4\pi P^{(0)}(r)r^{3}}{r[r-2Gm(r)]}~,
\end{equation}
where $P^{(0)}$ and $\varepsilon^{(0)}$ denote the equation of state (EoS)
describing the stellar matter. 
The quantity $m(r)$, defined as
\begin{eqnarray}
m(r)  =  4 \pi \int_0^r \varepsilon^{(0)}(r') r'^{2}  dr' \, ,
\end{eqnarray}
stands for the amount of gravitational mass contained inside a sphere
of radius $r$, with $r$ denoting the distance from the center of the
star. The star's total gravitational mass, $M$, is then given
\begin{equation} 
M = m(R) =4\pi \int_{0}^{R} \varepsilon^{(0)} (r^{\prime}) r^{\prime2}
dr^{\prime} \, ,
\end{equation}
where $R$ denotes the radius of the star defined by $P(r=R)=0$.
Physically, the TOV equation describes the balance of gravitational
and internal pressure forces at each radial distance inside the
star. Both forces exactly cancel each other inside a static stellar
configuration, as described by the TOV equation.

The TOV equation is solved numerically, for a given model for the EoS,
by choosing a value for the star's central density and then
integrating Eq.~(\ref{TOV}) out to a radial location where the
pressure becomes zero.  So for any fixed choice of the EoS, the stars
form a one-parameter sequence (parameter $\varepsilon^0_c$).  An
entire family of compact stars is obtained by solving the TOV equation
for a range of central densities which result in the mass-radius
relationship of compact stars. It is characterized by the existence of
a maximum mass star (several maximum mass stars if permitted by the
model chosen for the EoS).  The stars are stable against gravitational
collapse if they are on the stellar branch for which $\partial
M/\partial\varepsilon^{(0)}_c > 0$.  Stars on the stellar branch where
$\partial M/\partial\varepsilon^{(0)}_c < 0$ are unstable agains radial
oscillations and will therefore not exist stably in the universe.

Each stellar model has unique solutions for $m(r)$, $P^{(0)}(r)$ and
$\varepsilon^{(0)}(r)$ in terms of which the internal gravitational
field (the metric coefficients) is defined as
\begin{eqnarray}
\label{phs}
\lambda^{(0)}(r) & = &
-\ln[1-{2Gm(r)}/{r}]~,\label{phs-2}\\ \nu^{(0)}(r) & = &
-\lambda^{(0)}(R_{0})-2G\int_{r}^{R_{0}} \frac{m(r^{\prime})+4\pi
  P^{(0)}(r^{\prime})r^{\prime3}}{r^{\prime}[r^{\prime}-2Gm(r^{\prime})]}dr^{\prime}~.
\end{eqnarray}
The internal field solutions are smoothly connected to the external
field solutions at the stellar surface, $r=R.$ Once the internal
pressure profiles are derived from the solution of the TOV equations,
it is possible to compute other astrophysical quantities like baryonic
mass, and also make the next step in the perturbation approach to
define the moment of inertia and tidal deformabilities, which are of
very great observational interest.

Of particular interest for astrophysical scenarios (stellar
evolution) is the expression for the total baryon mass
\begin{eqnarray}
\frac{dN_{B}(r)}{dr} & = & 4\pi
r^{2}\left(1-\frac{2Gm(r)}{r}\right)^{-1/2}n(r),
\end{eqnarray}
where $n(r)$ is the baryon number density and $N_{B}(R)$ is the total
baryon number of the star.  This number is a characteristic conserved
quantity and is very important in discussions of evolutionary
scenarios of compact stars, see, e.g.,
\citet{Bejger:2016emu,Ayvazyan:2013cva,Chubarian:1999yn,Poghosyan:2000mr}.  The
functions of the spherically symmetric solution in Eqs.~(\ref{series})
and (\ref{euler1}) can be found from Eq.~(\ref{einstein}) and
Eq.~(\ref{euler}) in zeroth order of the $\Omega$-expansion.

\section{Tidal deformability of compact stars}

The tidal deformability (TD) is a measure of the shape deformation
property of the astrophysical object under the gravitational influence
of another nearby object. To determine it in the first order we need
to consider a modification of the space-time metric when the
distribution of matter of the star becomes elliptic.  According to the
symmetries of the metric coefficients introduced in
Eq.~(\ref{metric-1}) we have even orders $j=0,2, \ldots$ for the
diagonal elements\footnote{Notation corresponds to the work of
  \citet{Chubarian:1999yn}.}.  The the first correction corresponding
to small deformations (small values of $\beta$) one can consider terms
linear in $\beta$ or equivalently the $j=2$ perturbation approximation
to the spherically symmetric star.  We introduce some new notation and
work under the assumption that $f_{2}(r)=-\Phi_{2}(r)=A(r)$ like in
the expansion of the external solution, since
$\nu^{(0)}(r)=-\lambda^{(0)}(r)$.

The non diagonal term could be taken to be zero, because we consider
only the static case $q(r,\theta)=0$ (the parameter defining the
static deformation $\sqrt{\beta}$ does not change the sign under time
reversal $t\rightarrow-t$) and $U_{2}(r)=K(r)$, so that we have
\begin{eqnarray}
ds^{2} & = & e^{\lambda^{(0)}(r)}\left[1+\beta
  A(r)P_{2}(\theta)\right]dt^{2}\nonumber \\ & &
-e^{\nu^{(0)}(r)}\left[1-\beta A(r)P_{2}(\theta)\right]dr^{2}\nonumber
\\ & & -r^{2}\left[1-\beta
  K(r)P_{2}(\theta)\right]\left(d\theta^{2}+\sin^{2}\theta
d\varphi^{2}\right)~.
\end{eqnarray}

In this approximation, without loss of generality, one can set the
values of $f_{0}=\Phi_{0}=U_{0}=0$, because we neglect the
contribution of the deformation energy to the gravitational mass.
{The equations show that $K'(r)=A'(r)+{\nu^{(0)}}'(r)A(r)$ the prime symbol
denoting the derivative of those quantities with respect to $r$.}  
The functions $A(r)$ and $B(r)=dA/dr$ obey the differential equations
\begin{eqnarray}
\frac{dA(r)}{dr} & = & B(r);\\
\frac{dB(r)}{dr} & = & 2\left(1-2G\frac{m(r)}{r}\right)^{-1}\nonumber \\
 & \times & A(r)\left\{ -2\pi\left[5\varepsilon^{(0)}(r)+9P^{(0)}(r)+\frac{1}{c_{s}^{2}}\left(\varepsilon^{(0)}(r)+P^{(0)}(r)\right)\right]\phantom{\frac{3}{r^{2}}}\right.\nonumber \\
 & + & \frac{3}{r^{2}}+2\left(1-2G\frac{m(r)}{r}\right)^{-1}\left[G\left(\frac{m(r)}{r^{2}}+4\pi rP^{(0)}(r)\right)\right]^{2}\}\nonumber \\
 & + & \frac{2B(r)}{r}\left(1-2G\frac{m(r)}{r}\right)^{-1}\nonumber \\
 & \times & \left\{ -1+G\left[\frac{m(r)}{r}+2\pi r^{2}\left(\varepsilon^{(0)}(r)-P^{(0)}(r)\right)\right]\right\} .
\end{eqnarray}
Here, $c_{s}^{2}=dP/d\varepsilon$ is the square of the speed of sound,
which is equivalent to the knowledge of the equation of state. The
pressure profile provided by solving the TOV equations will complement
the above equations.

The system is to be integrated with the asymptotic behavior of metric
functions $A(r)=a_{0}r^{2}$ and $B(r)=2a_{0}r$ as $r\to0$, The $a_{0}$
is a constant that quantifies the deformation of the star which can be
taken arbitrary. This constant corresponds to the choice of $\beta$ as
an external parameter. Since it cancels in the expression for the Love
number and in all other quantities in consideration its value is not
important. Using  the solution on the surface at $r=R$ and the following
combination
\begin{equation}
y=\frac{R\,B(R)}{A(R)} \, ,
\end{equation}
it is possible to compute the $l=2$ Love
number~\citep{Hinderer:2007mb,Damour:2009vw,Binnington:2009bb,Yagi:2013awa,Hinderer:2009ca}:
\begin{eqnarray*}
k_{2} & = & \frac{8C^{5}}{5}(1-2C)^{2}[2+2C(y-1)-y]\\
 & \times & \bigg\{2C[6-3y+3C(5y-8)]\\
 & + & 4C^{3}[13-11y+C(3y-2)+2C^{2}(1+y)]\\
 & + & 3(1-2C)^{2}[2-y+2C(y-1)]\ln(1-2C)\bigg\}^{-1},
\end{eqnarray*}
where {$M/R$} in the expression $C=GM/R$ is the compactness of the star 
($2C$ is the ratio of gravitational radius to spherical radius).

The dimensionless tidal deformability parameter is defined as
$\Lambda=\lambda/M^{5}$, a quantity defined for small tidal
deformabilities. Here $\lambda$ is the TD of the star with a 
gravitational mass $M$, just as defined above. In addition, the love
number is related to TD and defined as
\begin{equation}
k_{2}=\frac{3}{2}\lambda R^{-5}.\label{k2def}
\end{equation}

In the investigations and observations of the process of neutron star
merging the TD $\Lambda$ is a key parameter characterising the
stiffness of equation of state of the stellar matter.

\subsection{Moment of inertia}\label{ssec:moi}
\label{gravity:iinertia}

The moment of inertia is one of the main
characteristics of the mechanical properties of the rotating body,
therefore one needs to define it also for the relativistic objects as
neutron star obeying the gravitational field contribution to the
rotational motion. 
The baryonic mass is an important quantity often
associated with explosive events, where it can be conserved while the
gravitational mass of the star suffers
modifications~\citep{Alvarez-Castillo:2015dqa,Bejger:2016emu}. 
The moment of inertia is related to the \textit{glitch} phenomenon,
which is a sudden spin-up in the general spin-down evolution of rotation frequencies,
observed for some pulsars, see~\citep{Haskell:2015jra} and references therein. 
Moreover, it is expected to be
measured in pulsar binaries, providing a strong constraint on the
compact star EoS~\citep{Lattimer:2006xb}.

In a very simplified way
  one can estimate the impact of relativistic effects on the moment of
  inertia from \citep{1994ApJ...424..846R}
\begin{equation}
  I\simeq\frac{C_{J}}{1+2GJ/R^{3}} \, ,
  \label{eq:moi_CJ}
\end{equation}
where $J$ denotes the total, conserved angular moment.  The quantity
$C_J$ is given by
\begin{equation}
 C_{J} = \frac{8\pi}{3}\int_{0}^{R}r^{4}
 \left(\varepsilon^{(0)}(r)+P^{(0)}(r)\right)\frac{1}{1-2Gm(r)/r}dr \, ,
\end{equation}
which can be readily computed since only the knowledge of spherically
symmetric quantities is required, which are easy to compute.

However, because of the deformation of the star due to rotation and the
impact of the gravitational field on the  rotational inertia
of the star, the moment of inertia becomes a function of the rotational
state, i.e., a function of the angular velocity or spin frequency of
the neutron star. To take all these effects into account in the defining
expression for the moment of inertia we will follow the steps of our
perturbation approach.

By definition, the angular momentum of the star in the case of stationary
rotation is a conserved quantity and can be expressed in invariant
form 
\begin{equation}
J=\int T_{\phi}^{t}\sqrt{-g}dV~,\label{moment}
\end{equation}
where $\sqrt{-g}dV$ is the invariant volume and $g=\det||g_{\mu\nu}||$.
For the case of slow rotation where the shape deformation of the rotating
star can be neglected and using the definition of the moment of inertia
$I_{0}(r)=J_{0}(r)/\Omega$ accumulated in the sphere with radius
$r$, we obtain from Eq.~(\ref{moment}) 
\begin{eqnarray*}
\frac{dI_{0}(r)}{dr}=\frac{8\pi}{3}r^{4}
     [\varepsilon^{(0)}(r)+P^{(0)}(r)]e^{(-\nu^{(0)}(r)
       +\lambda^{(0)}(r))/2}\frac{\bar{\omega}(r)}{\Omega}~.
\end{eqnarray*}

Here $\bar{\omega}$ the difference of the frame dragging frequency
$-\omega$ and the angular velocity $\Omega$.
In general relativity, due to the Lense-Thirring law,
rotational effects are described by 
\begin{equation}
\bar{\omega}\equiv\Omega+\omega(r,\theta).
\label{dromega}
\end{equation}

This expression is approximated from the exact expression of the energy
momentum tensor coefficient $T_{\phi}^{t}$ and the metric tensor in the
axial symmetric case.

We keep only two non-vanishing components of the 4-velocity 
\begin{eqnarray}
u^{\phi} & = & \Omega~u^{t}~,\nonumber \\ u^{t} & = &
1/\sqrt{e^{\nu}-r^{2}e^{\mu}\bar{\omega}^{2}  \sin^{2}\theta}~.
\end{eqnarray}
because we assume that the star due to high viscosity (ignoring the
super-fluid component of the matter) rotates stationarly as a solid
body with an angular velocity $\Omega$ that is independent of the
spatial coordinates. The time scales for changes in the angular velocity
which we will consider in our applications are well separated from
the relaxation times at which hydrodynamical equilibrium is established,
so that the assumption of a rigid rotator model is justified.

Now besides of central energy density $\varepsilon(0)$ of the star
configuration the angular velocity of the rotation $\Omega$ is an
additional parameter of the theory.

As a next step, going beyond the spherically symmetric case that
corresponds to the first-order approximation, we solve
Eq.~(\ref{rote}), where the unknown function is $q(r,\theta)$ which is
defined by Eq.~(\ref{intq}) and scaled such that it is independent of
the angular velocity. Using the static solutions Eqs.~(\ref{TOV}) and
(\ref{phs}), and the representation of $q(r,\theta)$ by the series of
the Legendre polynomials,
\begin{equation}
q(r,\theta)=\sum_{m=0}^{\infty}q_{m}(r)\frac{dP_{m+1}(\cos\theta)}{d\cos\theta}~,
\end{equation}
one can see that this series is truncated and only the coefficient
$q_{0}(r)$ is nontrivial, i.e., $q_{m}(r)=0$ for $m>0$ , see
\citep{Hartle:1967he,Chubarian:1999yn}.  Therefore one can write down
the equations for $\bar{\omega}(r)=\Omega(1+q_{0}(r)/{\Omega}_{0})$,
which is more suitable for the solution of the resulting equation in
first order
\begin{equation}
\frac{1}{r^{4}}\frac{d}{dr}\left[r^{4}j(r)
  \frac{d\bar{\omega}(r)}{dr}\right] +
\frac{4}{r}\frac{dj(r)}{dr}\bar{\omega}(r)=0~,
\label{Hartl}
\end{equation}
which corresponds to Ref.~\citep{Hartle:1967he}, where it was obtained
using a different representation of the metric. In this equation we
use the notation $j(r)\equiv e^{-(\nu^{(0)}(r)+\lambda^{(0)}(r))/2}$,
where $j(r)=1$ for $r>R_{0}$, i.e., outside of stellar configuration.

Using this equation one can reduce the second order differential equation
(\ref{Hartl}) to the first order one 
\begin{equation}
\frac{d\bar{\omega}(r)}{dr}=\frac{6GJ_{0}(r)}{r^{4}j(r)}.
\label{dmoment}
\end{equation}
and solve (\ref{Hartl}) as a coupled set of first order differential
equations, one for the moment of inertia (\ref{moment}) and the
other (\ref{dmoment}) for the frame dragging frequency $\bar{\omega}(r)$.

This system of equations is valid inside and outside the matter
distribution.  In the center of the configuration $I_{0}(0)=0$ and
$\bar{\omega}(0)=\bar{\omega}_{0}$.  The finite value
$\bar{\omega}_{0}$ has to be defined such that the dragging frequency
$\bar{\omega}(r)$ smoothly joins the outer solution
\begin{equation}
\bar{\omega}(r)=\Omega\left(1-\frac{2GI_{0}}{r^{3}}\right).\label{dgvac}
\end{equation}
at $r=R_{0}$, and approaches $\Omega$ in the limit $r\to\infty$.  In
the external solution (\ref{dgvac}) the constant $I_{0}=I_{0}(R_{0})$
is the total moment of inertia of the slowly rotating star and
$J_{0}=I_{0}\Omega$ is the corresponding angular momentum. In this
order of approximation, $I_{0}$ is a function of the central energy
density or the total baryon number only. This solution remains
connected to the spherically distributed matter and therefore does
not differ too much from our previous expression, which uses this
solution to incorporate the relativistic corrections.

However, to find the explicit dependence of the moment of inertia on
the angular velocity, one needs to take the second step and account
for the deformation of the stellar configuration, which, in the
framework of our scheme, is a second-order correction.

\subsection{Rotational deformation and moment of inertia}
\label{gravity:iinertia2}

To calculate these contributions and the internal structure of the
rotating star which is deformed due to centrifugal force one needs
to return to our perturbation description and take into the corrections
in the diagonal elements of the metric and the energy-momentum tensor, as
in the equations  above {with} the  parameter $\beta=(\Omega/\Omega_{0})^2$.

For a more detailed description of the solutions of the field
equations in the $\sim O(\Omega^{2})$ approximation we refer to the
works of \citet{Hartle:1967he} as well as
\citet{Chubarian:1999yn}. Since these equations have a complicated
form, here we will discuss only the qualitative meaning of the
physical quantities concerning the star's deformation and its moment
of inertia.

In $\Omega^{2}$-approximation the shape of the star is an ellipsoid,
and each of the equal-pressure (isobar) surfaces in the star is an
ellipsoid as well. All diagonal elements of the metric and
energy-momentum tensors could be represented as a series expansion in
Legendre polynomials, as we have already discussed in the previous
section where is has been noted that the only non vanishing solutions
obeying the continuity conditions on the surface with the external
solution of fields are those with $l=0,2$.

The deformation of the isobaric surfaces can be parameterised by the
deformation shifts $R(r,\theta)-r=\Delta(r,\theta)$ from the spherical
shape. It describes the deviation from the spherical distribution
as a function of radius $r$ for a fixed polar angle $\theta$
and is completely determined by 
\begin{equation}
R(r,\theta)=r+\left(\frac{\Omega}{\Omega_{0}} \right)^{2}
[\Delta_{0}(r)+\Delta_{2}(r)P_{2}(\cos\theta)],\label{def}
\end{equation}
since the expansion coefficients of the deformation $\Delta_{l}(r)$
are connected with the pressure corrections 
\begin{equation}
\Delta_{l}(r)=-\frac{p^{(l)}(r)}{dP^{(0)}(r)/dr}~.\label{ds}
\end{equation}
$l\in\{0,2\}$ is the polynomial index in the angular expansion
{
in Legendre polynomials,
analogous to Eq.~(\ref{Legendre-exp}). 
}
The function $R(R_{0},\theta)$ is the radius where $p(R(R_{0},\theta))=0.$
To avoid confusion, we denote the spherical radius as $R_{0}$, which
not anymore the actual radius, but rather $R(R_{0},\theta)$ which is
the distance of the stellar surface from the center of the
configuration at a polar angle $\theta$. In particular, we define the
equatorial radius as $R_{e}=R(R_{0},\theta=\pi/2)$, the polar radius
as $R_{p}=R(R_{0},\theta=0)$, and the eccentricity as
$\epsilon=\sqrt{1-(R_{p}/R_{e})^{2}}$, all three quantities
characterizing the deformed shape of the star.

Using this the same approach we write the correction to the moment of
inertia as $\Delta I(r)=I(r)-I_{0}(r)$ and represent it as a sum of several
different contributions, 
\begin{equation}
\Delta I=\Delta I_{{\rm Redist.}} + \Delta I_{{\rm Shape}}+\Delta
I_{{\rm Field}}+\Delta I_{{\rm Rotation}}~.
\end{equation}

Since these contributions are obtained from the exact expression of
angular momentum in the integral form the first three contributions
can also be expressed by integrals of the form 
\begin{equation}
\Delta I_{\alpha}=\int_{0}^{I_{0}(R_{0})}dI_{0}(r)[W_{0}^{(\alpha)}(r)
  - {W_{2}^{(\alpha)}(r)}/{5}]~,\label{di}
\end{equation}
where integration is taken from the angular averaged modifications
of the matter distribution, the shape of the configuration and the
gravitational fields, 
\begin{eqnarray}
W_{l}^{{\rm (Field)}}(r) & = & \left(\frac{\Omega}{\Omega_{0}}\right)^{2}\left\{ 2U_{l}(r)-[f_{l}(r)+\Phi_{l}(r)]/{2}\right\} ,\\
W_{l}^{{\rm (Shape)}}(r) & = & \left(\frac{\Omega}{\Omega_{0}}\right)^{2}\frac{d~\Delta_{l}(r)}{dr}~,\\
W_{l}^{{\rm (Redist.)}}(r) & = & \left(\frac{\Omega}{\Omega_{0}}\right)^{2}\frac{p_{l}(r)+\varepsilon_{l}(r)}{p^{(0)}(r)+\varepsilon^{(0)}(r)}~,
\end{eqnarray}
respectively. All quantities appearing have been determined from the
Eq.~(\ref{einstein}) in second order approximation. The contribution
of the change of the rotational energy to the moment of inertia is
given by
\begin{equation}
\label{eq:DI}
\Delta I_{{\rm Rotation}}=\frac{4}{5}\int_{0}^{I_{0}(R_{0})}dI_{0}(r)
\left[r^{2}\bar{\omega}^{2}(r)e^{-\nu_{0}(r)}\right]~.
\end{equation}
and includes the frame dragging contribution.

In the next sections of this chapter we will discuss results for the
moment of inertia along with stability conditions. We note that a
consistent discussion of the stability of rotating stars requires one
to take into account the contribution of the rotational energy to the
mass energy, as well as the corresponding corrections to the moment of
inertia.

\section{Models for the EoS with a strong phase transition }
\label{eos}

In this contribution, we focus on EoS models which describe a strong
phase transition in the sense that upon solving the TOV equations with
them compact star sequences are obtained which exhibit a third family
branch  in the mass-radius or mass-central (energy) density
diagram which is separated from the second family of neutron stars by
a sequence of unstable configurations.  The possibility of the very
existence of a third family of compact stars as a consequence of a
strong phase transition in dense nuclear matter, together with a
sufficient stiffening of the high-density matter that can be expressed
by a strong increase in the speed of sound (but not violating the
causality bound) has been discusses by Gerlach as early as 1968
\cite{Gerlach:1968zz}.  Let us note here that the very existence of such a
third family of compact stars is an effect of strong, general
relativistic gravity!  Namely, that the compactification which
accompanies the strong phase transition of the star leads to a
reduction of the gravitational mass of the hybrid star configuration
from which it only recovers (and thus escapes  gravitational
collapse) when after the transition the hybrid star core consists of
sufficiently stiff high-density matter.

Such EoS lead to mass-radius relationships for hybrid stars that have
been classified (D)isconnected or (B)oth in
Ref.~\cite{Alford:2013aca}. The "D" topology consists of a hadronic
and a hybrid star branch, both of which being gravitationally
disconnected from each other.  In contrast to this, the "B" topology
consists of a branch of stable hadronic stars followed by stable
hybrid stars, which are gravitationally disconnected from a second
branch of stable hybrid stars.  For the introduction of this
classification scheme, the constant-speed-of-sound (CSS) EoS was used
in \cite{Alford:2013aca} for describing the high-density matter, see
also Ref.~\cite{Zdunik:2012dj} for the justification of its validity
in the case of color superconducting quark matter EoS.  The first
demonstration that high-mass twin stars and thus a corresponding
high-mass third family sequence with $M_{\rm max} > 2~M_\odot$ were
possible, has been given in \cite{Alvarez-Castillo:2013cxa} The
intricacy of an equation of state describing a third family of stars
(with twin stars at high or low masses as a consequence) consists in
the fact that one needs, on the one hand, a sufficiently large jump in
energy density $\Delta \varepsilon$ and a relatively low critical
energy density $\varepsilon_c$ at the transition point to fulfil the
Seidov criterion (\ref{seidov}) for gravitational instability (i.e., a
stiff nuclear matter EoS has to be followed by a soft high-density
one), while on the other the high-density EoS needs to become
sufficiently stiff directly after the phase transition, without
violating the causality condition ($c_s^2<1$).  With the CSS
parametrization, these constraints could be fulfilled relatively
straightforwardly by dialling $c_s^2=1$ and adjusting a sufficiently
large value of $\Delta \varepsilon$ by hand.

The question arose whether a hybrid star EoS describing a third family
of compact stars with a maximum mass above $2~M_\odot$ could also be
obtained when applying the standard scheme of a two-phase approach
based on a realistic nuclear matter EoS and a microscopically
well-founded quark matter EoS, both joined, e.g., by a Maxwell
construction.  A positive answer was given already in 2013, when two
examples of this kind were presented in \cite{Blaschke:2013ana}, where
the excluded-volume corrected nuclear EoS APR and DD2 were joined with
a quark matter EoS based on the nonlocal NJL model approach
\cite{Blaschke:2007ri,Benic:2013eqa}, augmented with a density
dependent repulsive vector meanfield that was constructed by employing
a thermodynamically consistent interpolation scheme introduced in
Ref.~\cite{Blaschke:2013rma}.  Such an interpolation scheme, based on
the nonlocal, color superconducting NJL model of
\cite{Blaschke:2007ri}, but extended to address also a
density-dependent bag-pressure that facilitates a softening of the
quark matter EoS in the vicinity of the deconfinement transition, has
recently been developed in \cite{Alvarez-Castillo:2018pve}, guided by a
relativistic density functional (RDF) approach to quark matter
\cite{Kaltenborn:2017hus} with an effective confinement mechanism
according to the string-flip model
\cite{Horowitz:1985tx,Ropke:1986qs}.  This approach has been applied
very successfully to describe a whole class of hybrid EoS with a third
family branch fulfilling the modern compact star constraints
\cite{Ayriyan:2017nby,Alvarez-Castillo:2018pve}.  In the RDF approach
to the string-flip model of quark matter an essential element is the
ansatz for the nonlinear density functional resembling confinement and
embodying the aspect of in-medium screening of the string-type
confining interaction within an excluded volume mechanism. Another
nonlinearity term in the density functional embodies the stiffening of
quark matter at high densities in a similar way as it was obtained
from 8-quark interactions in the NJL model \cite{Benic:2014iaa} that
were an essential part of the description of high-mass twins in
\cite{Benic:2014jia}.  Both nonlinearities, due to (de)confinement and
high-density stiffening, are mimicked in a rather flexible way by the
twofold interpolation scheme suggested in
\cite{Alvarez-Castillo:2018pve} which can be reinterpreted as a
generalization of the nonlocal NJL model with
chemical-potential-dependent parameters.

Having discussed the successful approaches to construct EoS with a
strong phase transition which account for third family branches of
compact stars and can be recognized observationally by the mass twin
phenomenon, we would like to mention which ingredients are
indispensable for obtaining this feature and which approaches have
failed to obtain it.  An excellent illustration of the various
possibilities to join by interpolation hadronic and quark matter
phases which themselves have different characteristics of stiffness,
can be found in the recent review of
Ref.~\cite{Baym:2017whm}. However, despite being quite general, the
case of the third family branch and mass-twin compact stars could not
be captured!  The reason can be found elucidated in
Ref.~\cite{Alvarez-Castillo:2018pve}, where it is demonstrated for a
representative set of hadronic as well as quark matter EoS of varying
degree of stiffness, that either a phase transition in the relevant domain
of densities is entirely absent or results in a hybrid star branch
that is directly connected to the hadronic brach which therefore does not form
a third family of stars.  The generated patterns are very similar to the results of
Ref.~\cite{Orsaria:2013hna} which also uses the nonlocal NJL model for
describing the quark phase of matter.  The clue to obtaining third
family sequences within microscopically motivated studies is a subtle
softening followed by a stiffening of quark matter that
can be realized employing the thermodynamically consistent
interpolation procedure between different parametrizations of the same
quark EoS (e.g., varying the vector meson coupling strength) before
applying a Maxwell-, Gibbs-, or pasta phase transition construction.

We have described here the state-of-the-art modeling of EoS with a
strong phase transition that are based on microscopic models of
high-density (quark) matter. Besides these, there is a large number of
simple EoS parametrizations that are also in use for discussing third
family sequences fulfilling the constraint of a high maximum mass of
the order of $2~M_\odot$.  These are basically the classes of CSS
based models and multi-polytrope approaches.  Without attempting
completeness, we would like to mention
Refs.~\cite{Alford:2015dpa,Alford:2015gna,Christian:2017jni,Alford:2017qgh,Paschalidis:2017qmb,Christian:2018jyd,Montana:2018bkb,Han:2018mtj}
from the class of CSS models. A particularly interesting work is the
extension by Alford and Sedrakian~\cite{Alford:2017qgh}, who
demonstrated that also a fourth family of hybrid stars can be obtained
and besides mass-twins there are also mass-triples possible.  The
multi-polytrope EoS have been a workhorse for numerical relativity
studies of astrophysical scenarios since many years.  The approach to
constrain the multi-polytrope parameters from observations of masses
and radii as introduced by Read et al. \cite{Read:2008iy} has then
been developed further with great resonance in the community by
\citet{Hebeler:2010jx,Hebeler:2013nza}. While in
\citet{Read:2008iy} one already finds a one parameter set describing
high-mass twin stars that have not yet become a matter of interest in
the community, the third family branch in the multi-polytrope
aproaches has been mainly overlooked (see, e.g.,
\citet{Raithel:2016bux,Miller:2019nzo}), but  was digged up in
Ref.~\cite{Alvarez-Castillo:2017qki} where it was found that one
should have at least a four-polytrope ansatz and suitably chosen
densities for the matching of the polytrope pieces of the EoS, see
also \citet{Annala:2017llu,Paschalidis:2017qmb,2018EPJWC.17120004H}.

An important issue when discussing strong first-order phase
transitions is the appearance of structures of finite size, like
bubbles and droplets in the boiling/condensation transitions of the
water-vapour transformations. In general, different shapes in the new
phase are possible like spherical, cylindrical and planar structures,
which have been dubbed "pasta structures". Their size and
thermodynamical favorability depends on the surface tension between
the subphases and the effects of the Coulomb interaction, including
screening.  The resulting pressure in the mixed phase is then no
longer constant as in the Maxwell construction case, but also not as
dramatically changing when the surface tension is neglected 
{
\cite{Glendenning:1992vb}.
For details concerning the quark-hadron transition pasta phases under 
neutron star constraints see, e.g., Refs.~\cite{Na:2012a,Yasutake:2014oxa,Spinella:2016EJP}.
}
It is interesting to note that a simple one-parametric parabolic
approximation of the pressure versus chemical potential dependence can
give a satisfactory agreement with a full pasta phase calculation and
that the single parameter can be directly related to the surface
tension \cite{Maslov:2018ghi}. It could be demonstrated that the third
family feature of an EoS with strong phase transition is rather robust
against pasta phase effects, see \citet{Ayriyan:2017nby}.

In the present work, we will use the multi-polytrope approach to
the EoS describing a third family of compact stars and also discuss
the effect of mimicking the pasta structures in the mixed phase by a
polynomial interpolation. The results shall not depend qualitatively
on these simplifying assumptions but be of rather general nature and
also applicable to more realistic types of EoS as discussed above.

\subsection{Multi-polytrope approach to the EoS}
\label{eos:polytrope}

In this section we present an EoS model that features a strong
first-order phase transition from hadron to quark matter. They are
labeled ``ACB'' following~\citep{Paschalidis:2017qmb} and consist of a
piecewise polytropic
representation~\citep{Read:2008iy,Hebeler:2013nza,Raithel:2016bux} of
the EoS at supersaturation densities ($n_{1}<n<n_{5}\gg n_{0}$):
\begin{eqnarray}
P(n)=\kappa_{i}(n/n_{0})^{\Gamma_{i}},\ n_{i}<n<n_{i+1},\ i=1\dots4,  \label{polytrope}
\end{eqnarray}
where $\Gamma_{i}$ is the polytropic index in each of the density
regions labeled by $i=1\dots4$. We fix $\Gamma_1$ such that
a stiff nucleonic EoS provided in~\citep{Hebeler:2013nza} can be
described.  The second polytrope shall correspond to a first-order
phase transition therefore in this region the pressure must be
constant, given by $P_{c}=\kappa_{2}$ ($\Gamma_{2}=0$). The remaining
polytropes, in regions 3 and 4, that lie above the phase transition
shall correspond to high-density matter, like stiff quark matter.

In order to compute the remaining thermodynamic variables of the
EoS, we utilize the formulae given in the Appendix of
Ref.~\citep{Zdunik:2005kh},
\begin{eqnarray}
P(n) & = & n^{2}\frac{d(\varepsilon(n)/n)}{dn},\\ \varepsilon(n)/n & =
& \int dn\,\frac{P(n)}{n^{2}}=\frac{1}{n_{0}^{\Gamma_i}}\int dn\,\kappa
n^{\Gamma_i-2}
=
\frac{1}{n_{0}^{\Gamma_i}}\frac{\kappa\,n^{\Gamma_i-1}}{\Gamma_i-1}+C,\\ \mu(n)
& = &
\frac{P(n)+\varepsilon(n)}{n}=\frac{1}{n_{0}^{\Gamma_i}}\frac{\kappa\,
  \Gamma_i}{\Gamma_i-1}n^{\Gamma_i-1}+m_{0},\label{5}
\end{eqnarray}
where we fix the integration constant $C$ by the condition that
$\varepsilon(n\to0)=m_{0}\,n$.  Now we can invert the above
expressions to obtain
\begin{eqnarray}
n(\mu) & = &
\left[n_{0}^{\Gamma_i}(\mu-m_{0})\frac{\Gamma_i-1}{\kappa\Gamma_i}\right]^{1/(\Gamma_i-1)},
\end{eqnarray}
so that the chemical potential dependent pressure for the polytrope
EoS (\ref{polytrope}) can be written as 
\begin{eqnarray}
P(\mu) & = & \kappa\left[n_{0}^{\Gamma_i}(\mu-m_{0})\frac{\Gamma_i-1}
  {\kappa\Gamma_i}\right]^{\Gamma_i/(\Gamma_i-1)}~.\label{P-mu}
\end{eqnarray}
The above form of the pressure (\ref{P-mu}) is suitable to perform a
Maxwell construction of a first-order phase transition between the
hadron and quark phases. The model parameters for the mass twin cases
that we consider in this work are given in table~\ref{param-45}.

\begin{table}[!htb]
\centering \tbl{ EOS models ACB4 and ACB5~\citep{Paschalidis:2017qmb}.
  The parameters are defined in Eq.~(\ref{polytrope}) in the main text.
  The first polytrope ($i=1$) describes the nuclear EoS at
  supersaturation densities, the second polytrope ($i=2$) corresponds
  to a first-order phase transition with a constant pressure $P_{c}$
  for densities between $n_{2}$ and $n_{3}$. The remaining polytropes
  lie in regions 3 and 4, i.e., above the phase transition and
  correspond to high-density matter, e.g., quark matter. The last
  column shows the maximum masses $M_{{\rm max}}$ on the hadronic
  (hybrid) branch corresponding to region 1 (4). The minimal mass
  $M_{{\rm min}}$ on the hybrid branch is shown for region 3.} {%
\begin{tabular}{c|c|cccc|c}
\hline 
 &  & $\Gamma_{i}$  & $\kappa_{i}$  & $n_{i}$  & $m_{0,i}$  & $M_{{\rm max/min}}$\tabularnewline
ACB  & i  &  & {[}MeV/fm$^{3}${]}  & {[}1/fm$^{3}${]}  & [MeV]  & $[M_{\odot}${]}\tabularnewline
\hline 
4  & 1  & 4.921  & 2.1680  & 0.1650  & 939.56  & 2.01 \tabularnewline
 & 2  & 0.0  & 63.178  & 0.3174  & 939.56  & -- \tabularnewline
 & 3  & 4.000  & 0.5075  & 0.5344  & 1031.2  & 1.96 \tabularnewline
 & 4  & 2.800  & 3.2401  & 0.7500  & 958.55  & 2.11 \tabularnewline
\hline 
5  & 1  & 4.777  & 2.1986  & 0.1650  & 939.56  & 1.40 \tabularnewline
 & 2  & 0.0  & 33.969  & 0.2838  & 939.56  & -- \tabularnewline
 & 3  & 4.000  & 0.4373  & 0.4750  & 995.03  & 1.39 \tabularnewline
 & 4  & 2.800  & 2.7919  & 0.7500  & 932.48  & 2.00 \tabularnewline
\hline 
\hline 
\end{tabular}
\label{param-45} 
} 
\end{table}

\begin{figure}[!htph]
\centering 
\includegraphics[width=\textwidth]{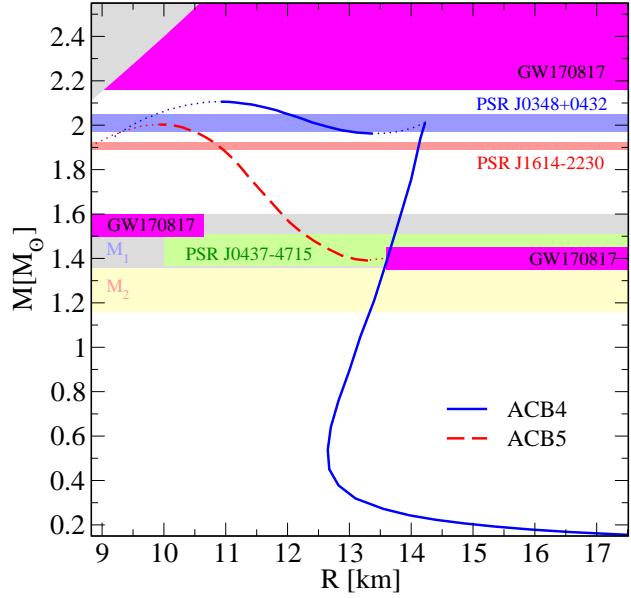} 
\caption{Mass-radius relationship. The two curves correspond to the
  compact star sequences resulting after integration of the TOV
  equations for the ACB4 and ACB5 EoS. The blue, red and yellow
  regions correspond to mass measurements of the PSR J0348+0432, PSR
  J1614-2230 and PSR J0437-4715 pulsars, respectively. The latter is
  the target of the NICER detector that shall provide a measurement
  for its radius~\citep{Arzoumanian:2009qn}.  The areas labeled M1 and
  M2 correspond to the mass estimates of the compact stars that merged
  in the GW170817 event that was detected through gravitational
  radiation~\citep{TheLIGOScientific:2017qsa}. The magenta marked
  areas labeled GW170817 are excluded by the GW170817
  event~\citep{Bauswein:2017vtn}.  An upper limit on the maximum mass
  of nonrotating compact stars of $2.16 M_{\odot}$ has been estimated
  in \cite{Rezzolla:2017aly}.  The region excluded by this estimate is
  shown by the magenta area, too.  This limit will be reconsidered
  again later in light of the material presented in this work.  The
  grey area in the upper left corner corresponds to a forbidden region
  where causality is violated.
\label{MvsR_ACB4_ACB5}}
\end{figure}

The ACB4 EoS features a first order phase transition at a rather high
nucleon number density value, $n_{2}=0.3174$~fm$^{-3}$ that produces
an instability in a $2.0 M_{\odot}$ neutron star, providing an example
of the high-mass twins phenomenon.  On the contrary, the ACB5 EoS
presents a phase transition that occurs at the lower value of
$n_{2}=0.284~\mathrm{fm}^{-3}$.  In that case the instability occurs
for stars with $1.4 M_{\odot}$, providing a scenario for low-mass twin
stars.  This low density value for the phase transition of about two
times saturation density, is particularly feasible in neutron star
matter, where the effect of the isospin asymmetry manifests in the so
called asymmetry energy which stiffens the EoS with respect to the
symmetric case, equal number of protons and neutrons in hadronic
matter. Figure~\ref{MvsR_ACB4_ACB5} shows the resulting mass-relation
curves for these two EoS featuring mass twins together with
measurements and constraints regions.

\begin{figure}[!htph]
\centering 
\includegraphics[width=0.8\textwidth]{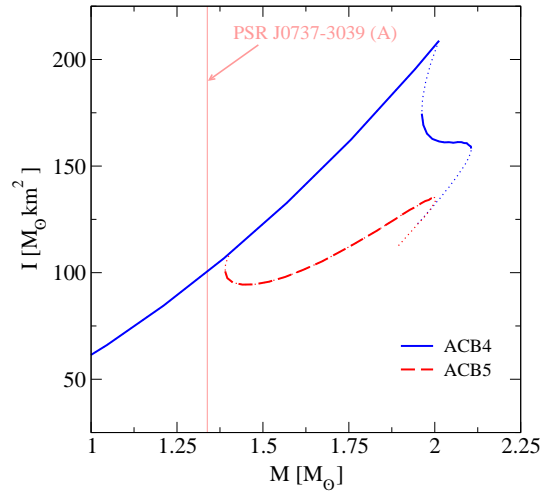} 
\caption{Moment of inertia as a function of gravitational mass of the
  star for the two EoS cases ACB4 and ACB5. For an orientation, we
  indicate the precisely determined mass of the star PSR J0737-3039
  (A), for which a measurement of the moment of inertia will become
  possible soon.  This will provide further constraints on the EoS of
  dense matter and astrophysical scenarios involving compact stars.
\label{IvsM-ACB4_ACB5}}
\end{figure}

\subsection{EoS including mixed phase effects (pasta phases)}
\label{eos:pasta}

In this section we introduce a mixed phase approach to mimic pasta
structures in regions of both the hadronic and quark EoS around the
Maxwell critical point ($\mu_{c},P_{c}$). The method used is the
replacement interpolation method (RIM)~\citep{Abgaryan:2018gqp}
that consists of replacing the EoS in the aforementioned domain by
a polynomial function:
\begin{equation}
  P_{M}\left(\mu\right) =
  \sum_{q=1}^{N}\alpha_{q}\left(\mu-\mu_{c}\right)^{q}
  +\left(1+\Delta_{P}\right)P_{c} \, ,
\label{mph_general}
\end{equation}
with $\Delta_{P}$ as free parameter that adds pressure to the mixed
phase at $\mu_{c}$.  Generally, all parametrizations of the type shown in
Eq.~(\ref{mph_general}) for the mixed phase pressure are even order
($N=2k$, k=1, 2, \dots) polynomials which we refer to as $G_k$.  In
order to smoothly match the EoS at $\mu_{H}$ and $\mu_{Q}$ up to the
$k$-th derivative of the pressure the following conditions shall be
fulfilled
\begin{eqnarray}
  P_{H}\left(\mu_{H}\right)&=&P_{M}\left(\mu_{H}\right) \, , \\
  P_{Q}\left(\mu_{Q}\right)&=&P_{M}\left(\mu_{Q}\right)\, ,\\
  \frac{\partial^{k}}{\partial\mu^{k}}P_{H}\left(\mu_{H}\right)&=&
  \frac{\partial^{k}}{\partial\mu^{k}}P_{M}\left(\mu_{H}\right) \, , \\
  \frac{\partial^{k}}{\partial\mu^{k}}P_{Q}\left(\mu_{Q}\right)&=&
  \frac{\partial^{k}}{\partial\mu^{k}}P_{M}\left(\mu_{Q}\right) \, ,
\label{press}
\end{eqnarray}
with $\alpha_{q}$, $\mu_{H}$ and $\mu_{Q}$ being determined by the
above system of equations.

\begin{figure*}[!htb]
\begin{centering}
\includegraphics[width=0.7\textwidth]{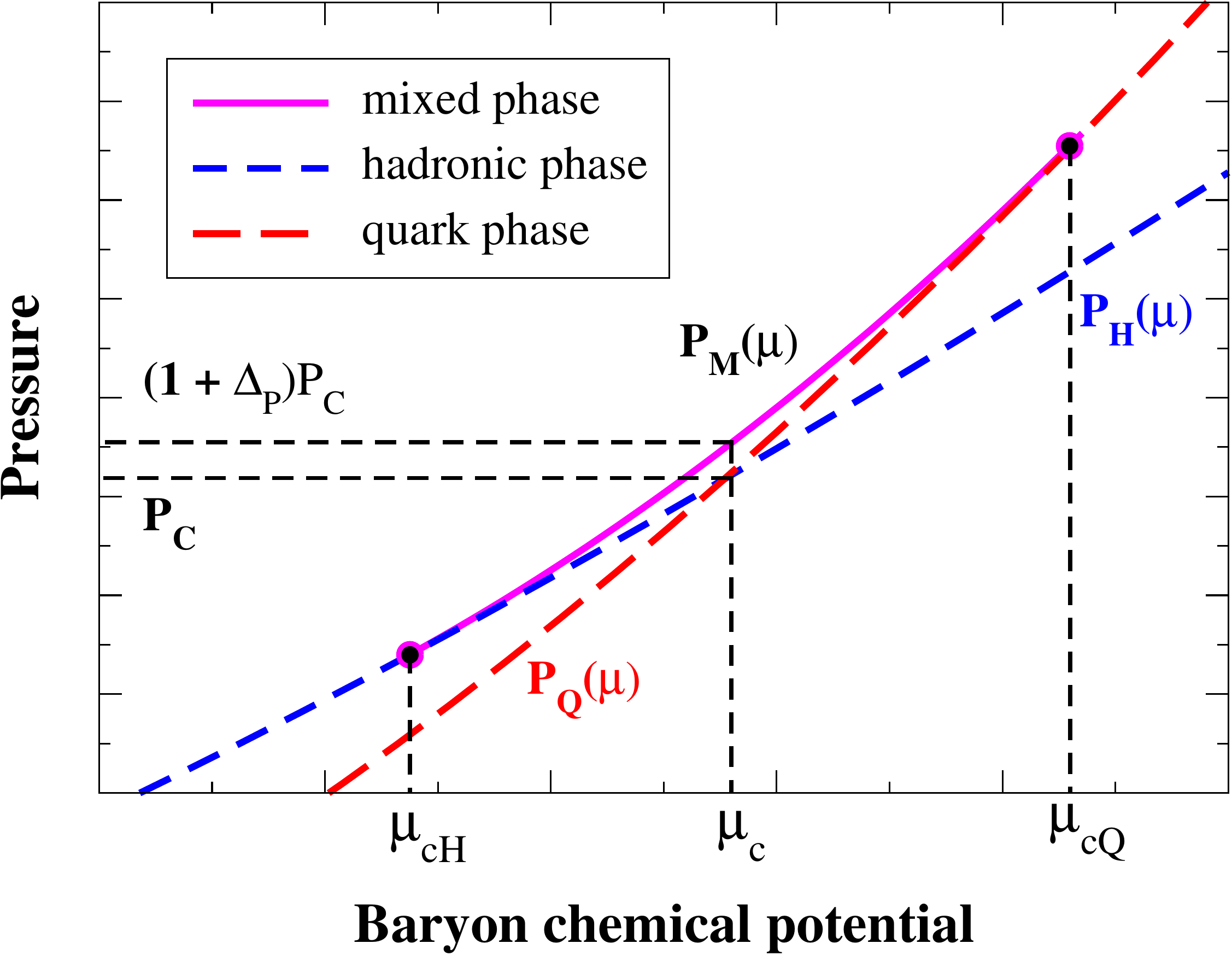} 
\par\end{centering}
 \caption{\label{RIM} Replacement interpolation function $P_{M}(\mu)$
   obtained from the mixed phase constructions around the Maxwell
   construction.  The resulting EoS connects the three points
   $P_{H}(\mu_{H})$, $P_{c}+\Delta P=P_{c}(1+\Delta_{P})$,
   and $P_{Q}(\mu_{Q})$. Figure taken from~\citep{Abgaryan:2018gqp}.}
\end{figure*}

For the sake of simplicity we employ the parabolic model $G_{1}$ of
the RIA as introduced in~\citep{Ayriyan:2017tvl,Ayriyan:2017nby}:
\begin{equation}
P_{M}\left(\mu\right)=\alpha_{2}\left(\mu-\mu_{c}\right)^{2} +
\alpha_{1}\left(\mu-\mu_{c}\right)+\left(1+\Delta_{P}\right)P_{c} \, ,
\end{equation}
where the parameters $\alpha_{1}$, $\alpha_{2}$, $\mu_{H}$ and
$\mu_{Q}$ are to be determined as described above, from the continuity
conditions at the Maxwell construction critical point: 
\begin{eqnarray}
P_{H}\left(\mu_{H}\right)&=&P_{M}\left(\mu_{H}\right) \, , \\
P_{Q}\left(\mu_{Q}\right)&=&P_{M}\left(\mu_{Q}\right)\, , \\
n_{H}\left(\mu_{H}\right)&=&n_{M}\left(\mu_{H}\right)\, , \\
n_{Q}\left(\mu_{Q}\right)&=&n_{M}\left(\mu_{Q}\right).
\end{eqnarray}
Figure~\ref{RIM} shows a schematic representation of the RIM method
based on the Maxwell construction between the hadronic and quark EoS.

\begin{figure*}[!htb]
\begin{centering}
$\begin{array}{cc}
\label{EoS_Pasta}
\includegraphics[width=0.48\textwidth]{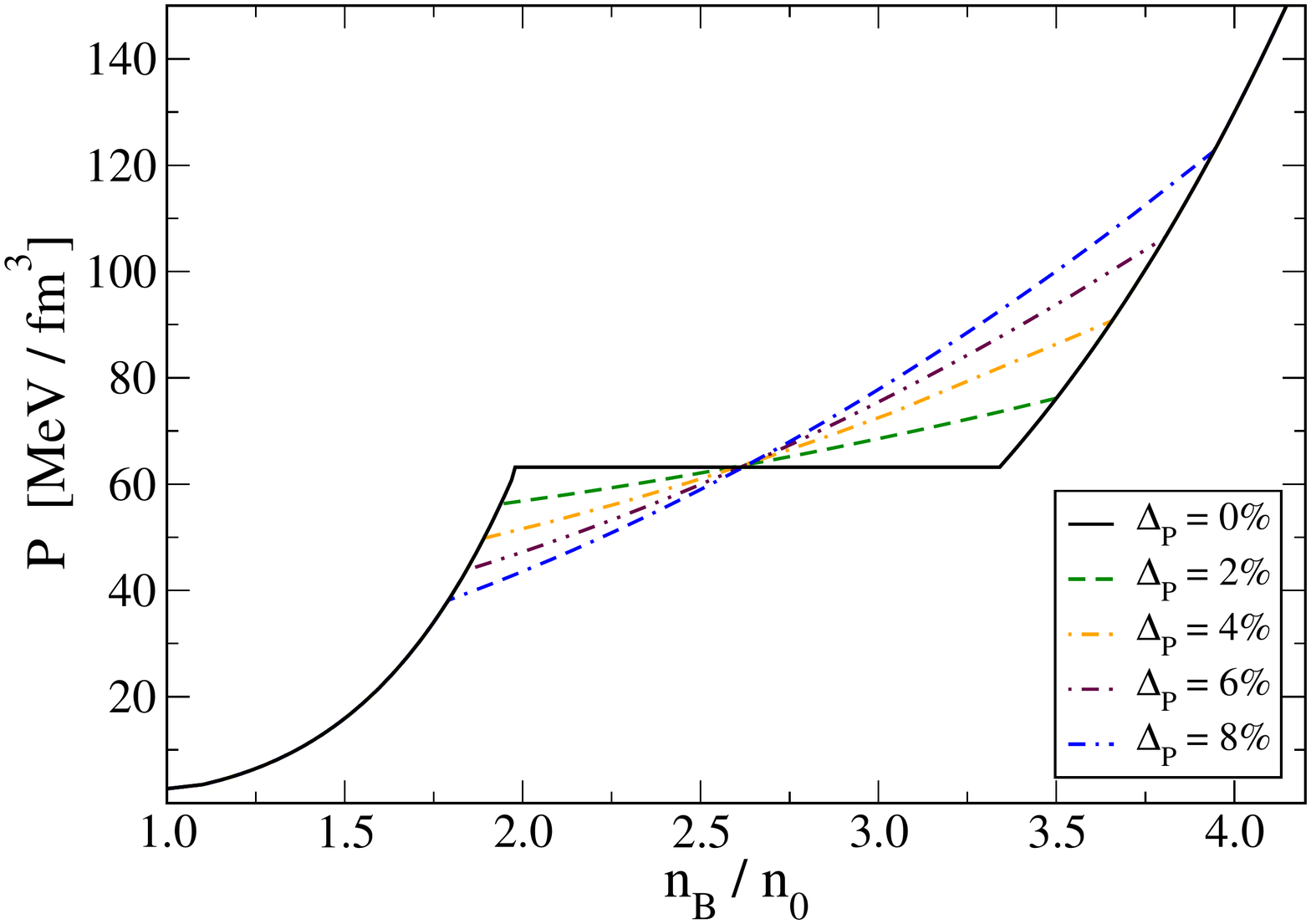} 
& \hspace{0cm}
\includegraphics[width=0.48\textwidth]{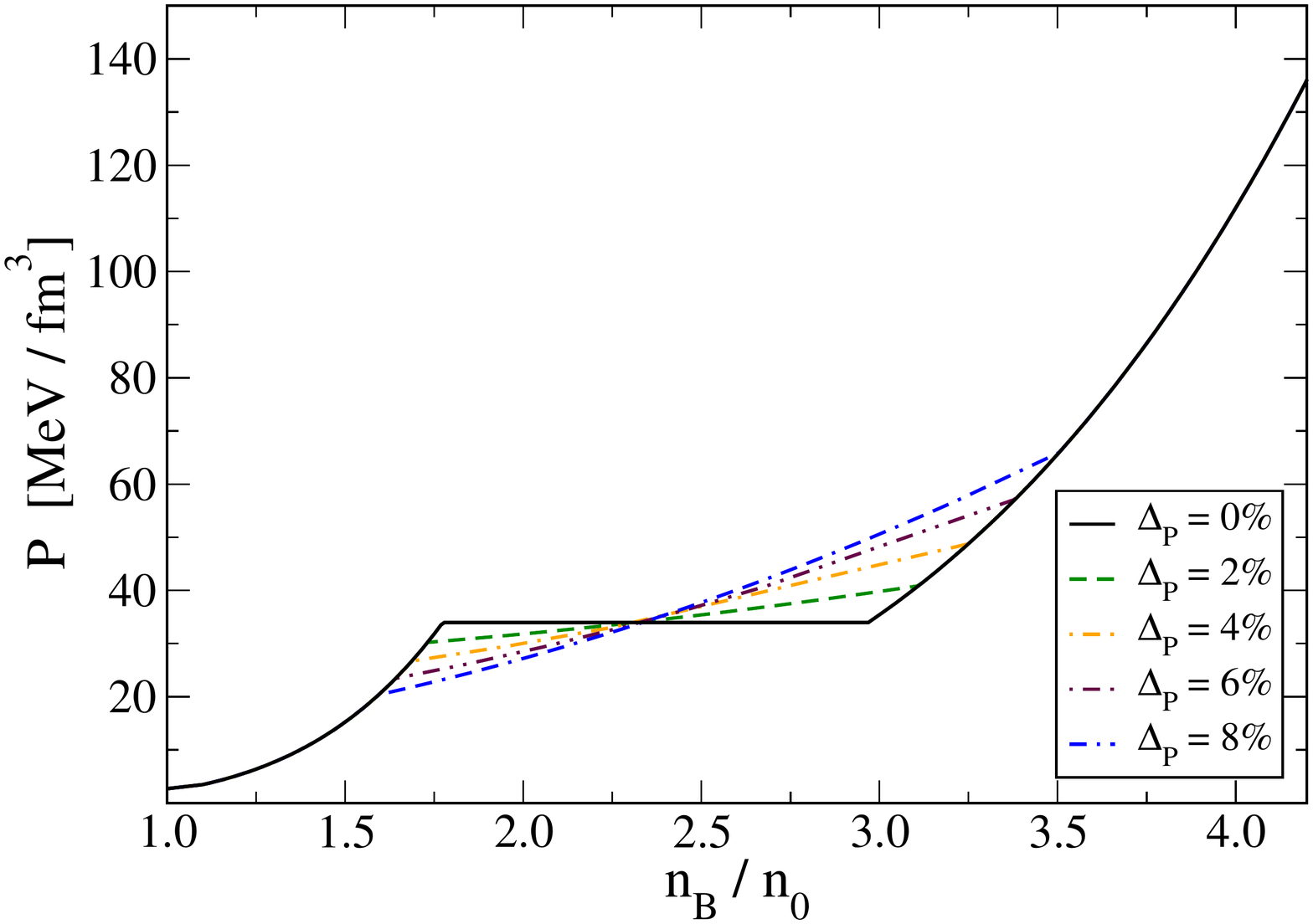}
\end{array}$ 
\par\end{centering}
\caption{Mixed phase mass twins equations of state for high mass NS
  onset (left) and low mass NS onset (right). The horizontal plateau
  at the phase transition corresponds to the Maxwell construction
  case. As the $\Delta_P$ parameter is increased successivley, the
  plateau gives way to straight lines with increasing slope values.}
\end{figure*}

\begin{figure*}[!bhtp]
\begin{centering}
$\begin{array}{cc}
\label{cs2_Pasta}
\includegraphics[width=0.48\textwidth]{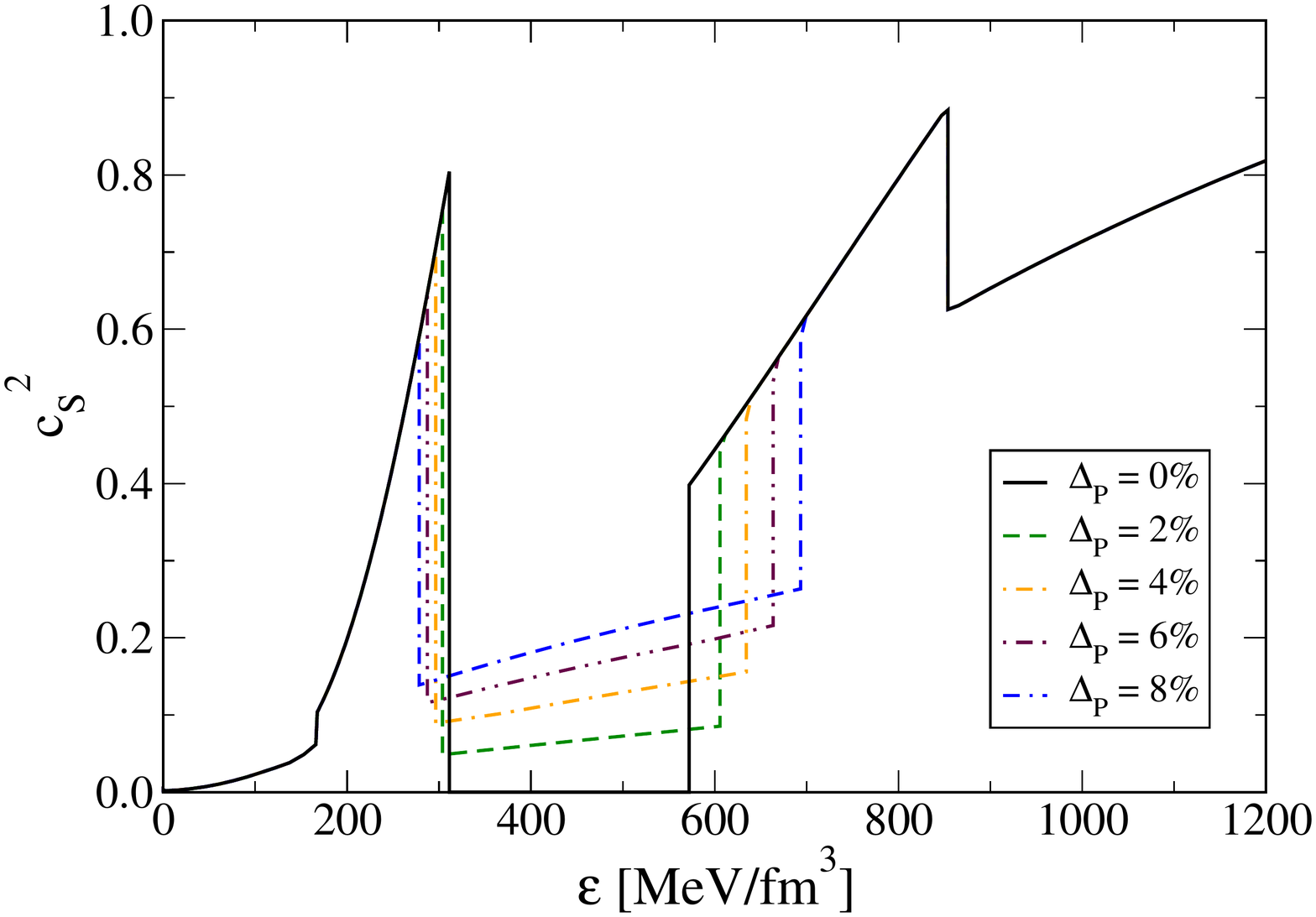} 
& \hspace{0cm}
\includegraphics[width=0.48\textwidth]{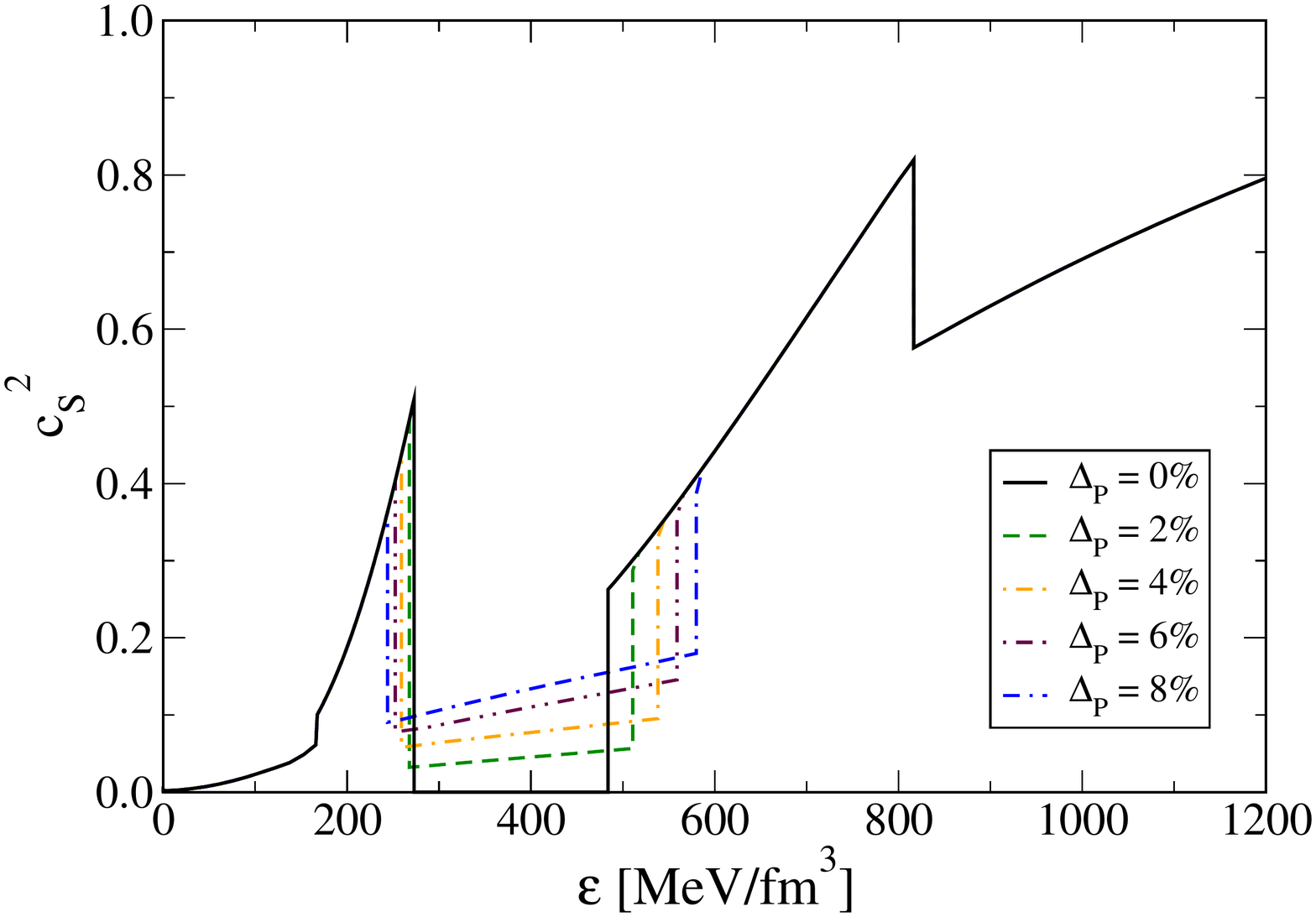}
\end{array}$ 
\par\end{centering}
\caption{Squared speed of sound as a function of the energy density
  for the mixed phase mass twins equations of state with high mass HS
  onset (ACB4, left panel) and low mass HS onset (ACB5, right
  panel). }
\end{figure*}

In addition, figure~\ref{EoS_Pasta} shows the mixed phase equations
of state for both low and high mass twins. The effect of the mimicked
geometrical structures is quantified by the $\Delta_{P}$ parameter.
It is evident that the order of the $G$ function will result in
whether or not there are discontinuities for the derivatives of the
$P_{M}$ function. For instance, the square of the speed of sound,
$c_{s}^{2}$, is proportional to the second derivative of $G$ with
respect to $\mu$, see figure~\ref{cs2_Pasta}.

The result is that $G_{1}$ presents  a clear discontinuity in the
speed of sound at $\varepsilon_{c}$ and
$\varepsilon_{c}+\Delta\varepsilon$, whereas in between i.e., in the
latent heat region, the speed of sound slightly increases.  On the
contrary, the construction $G_{2}$ allows for a continuous speed of
sound, however it is not smoothly connected at $\varepsilon_{c}$ and
$\varepsilon_{c}+\Delta\varepsilon$. Only $G_{3}$ is capable of
joining smoothly the speed of sound between the hadron and quark EoS
at the critical points.

\section{Results}

\subsection{TOV solutions for mixed phase models}
\label{results:tov}

In figure \ref{L1L2_Pasta} we show the results of the mass-radius diagram as a solution of the TOV equations for the equations of state  ACB4 (left panel) which exhibits high mass twin stars and ACB5 (right panel) which describes low mass twins, depending on the value of 
the mixed phase parameter $\Delta_P$.
In the insets we give a magnified view on the region of the maximum mass of the hadronic branch of the sequence, where the dotted lines indicate the unstable solutions that qualify the corresponding EoS as one with a third family. 
We can read off to the accuracy of the given 1\% steps what the critical value for $\Delta_P$ is when the disconnected second and third families would merge to a connected hybrid star sequence.

\begin{figure*}[!ht]
\begin{centering}
$\begin{array}{cc}
\label{fig:MR_ACB4-ACB5_Pasta}
\includegraphics[width=0.48\textwidth]{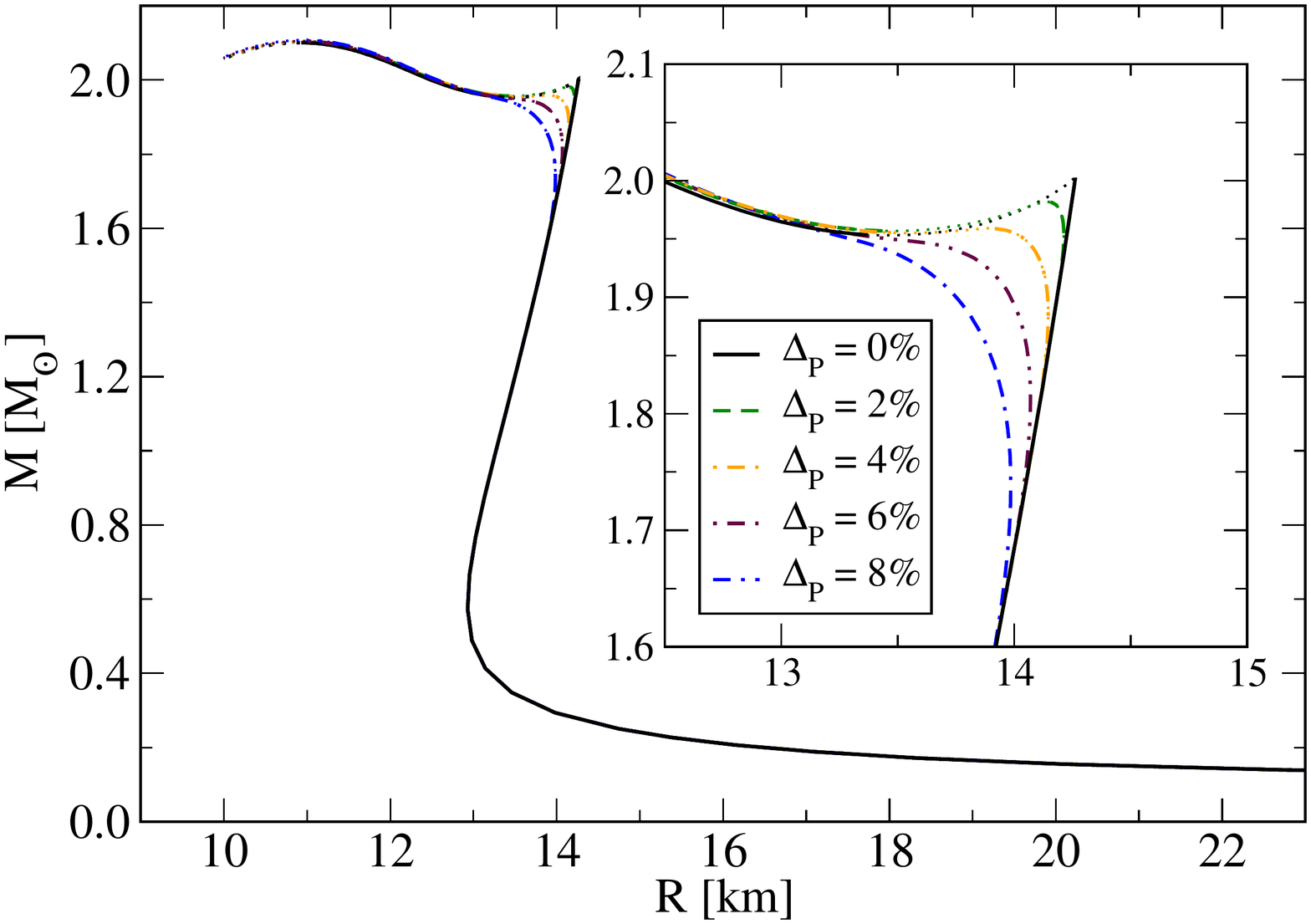} & \hspace{0cm}
\includegraphics[width=0.48\textwidth]{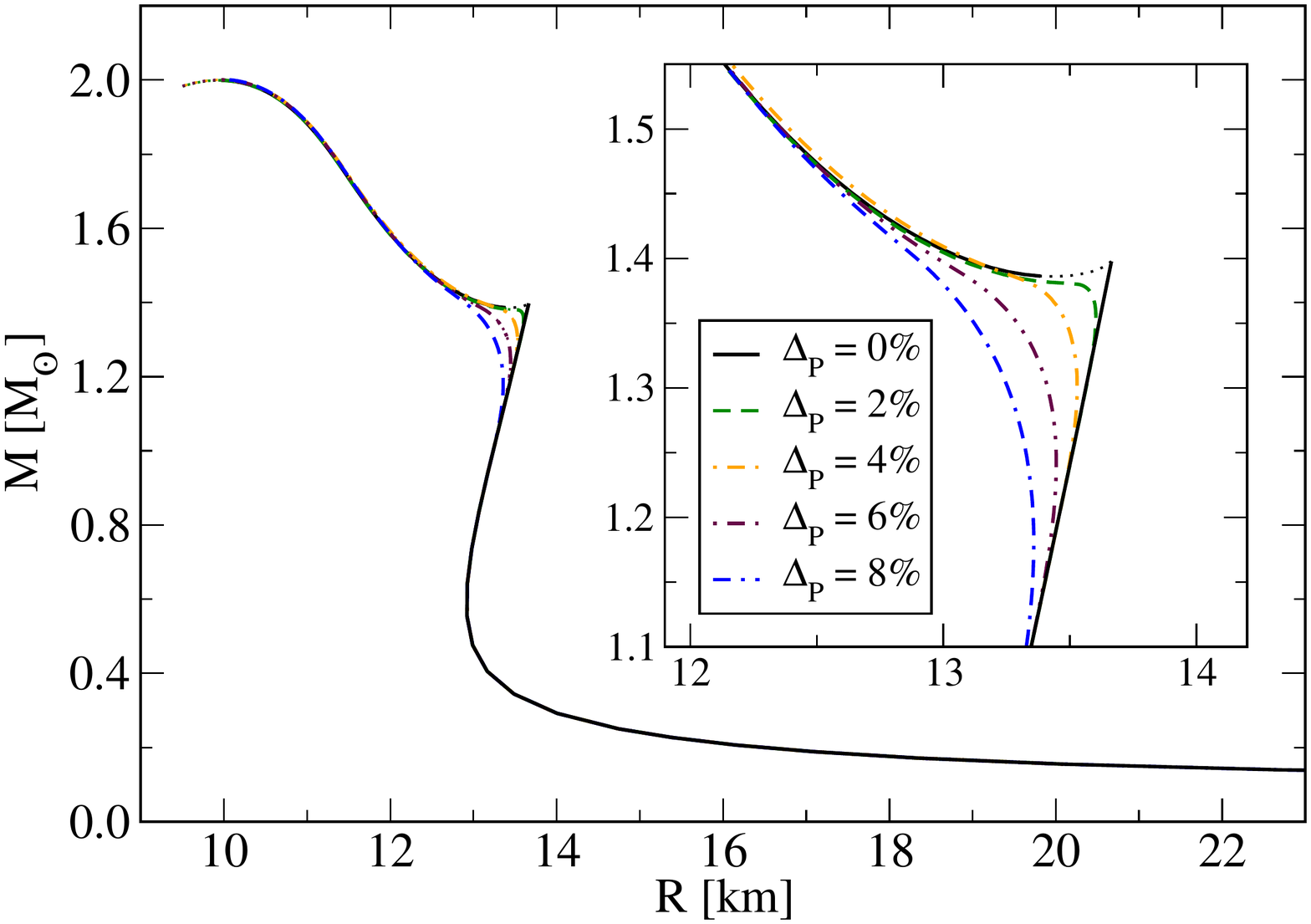}
\end{array}$ 
\par\end{centering}
\caption{Mass-radius diagram as a solution of the TOV equations for the equations of state ACB4 (left panel)
which exhibits high mass twin stars and ACB5 (right panel) which describes low mass twins, depending on the value of the mixed phase parameter $\Delta_P$. 
}
\end{figure*}

While for the case of ACB4 the variation of  $\Delta_P$ does not affect the mass-radius diagram in the mass region of the compact star merger GW170817, the corresponding variation for ACB5 leads to strong effects in that mass region. We therefore consider the tidal deformability in both cases in the next subsection.

\subsection{Tidal deformability predictions}
\label{results:lambda}

\begin{figure*}[!ht]
\begin{centering}
$\begin{array}{cc}
\label{L1L2_Pasta}\includegraphics[width=0.48\textwidth]{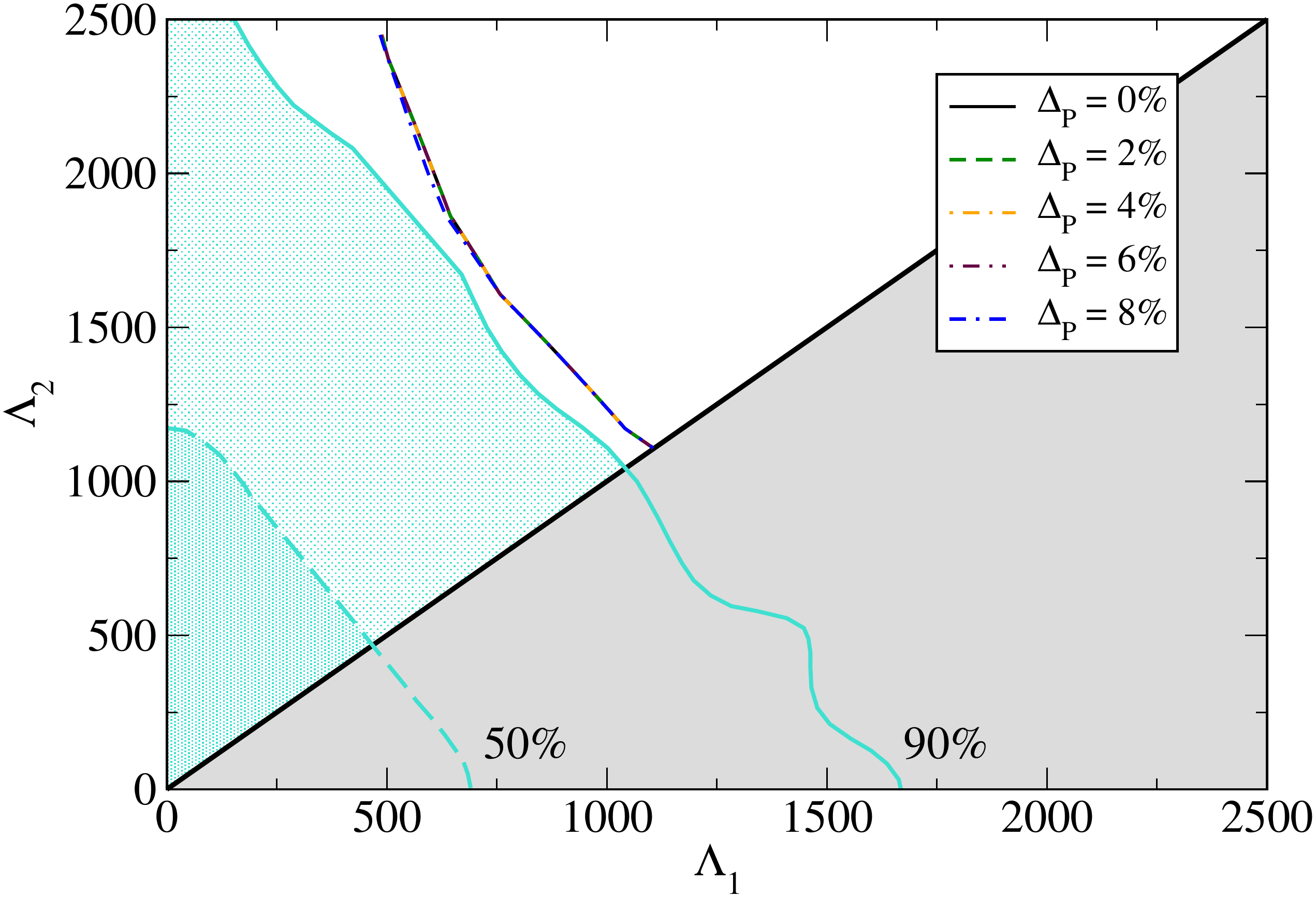} & \hspace{0cm}\includegraphics[width=0.48\textwidth]{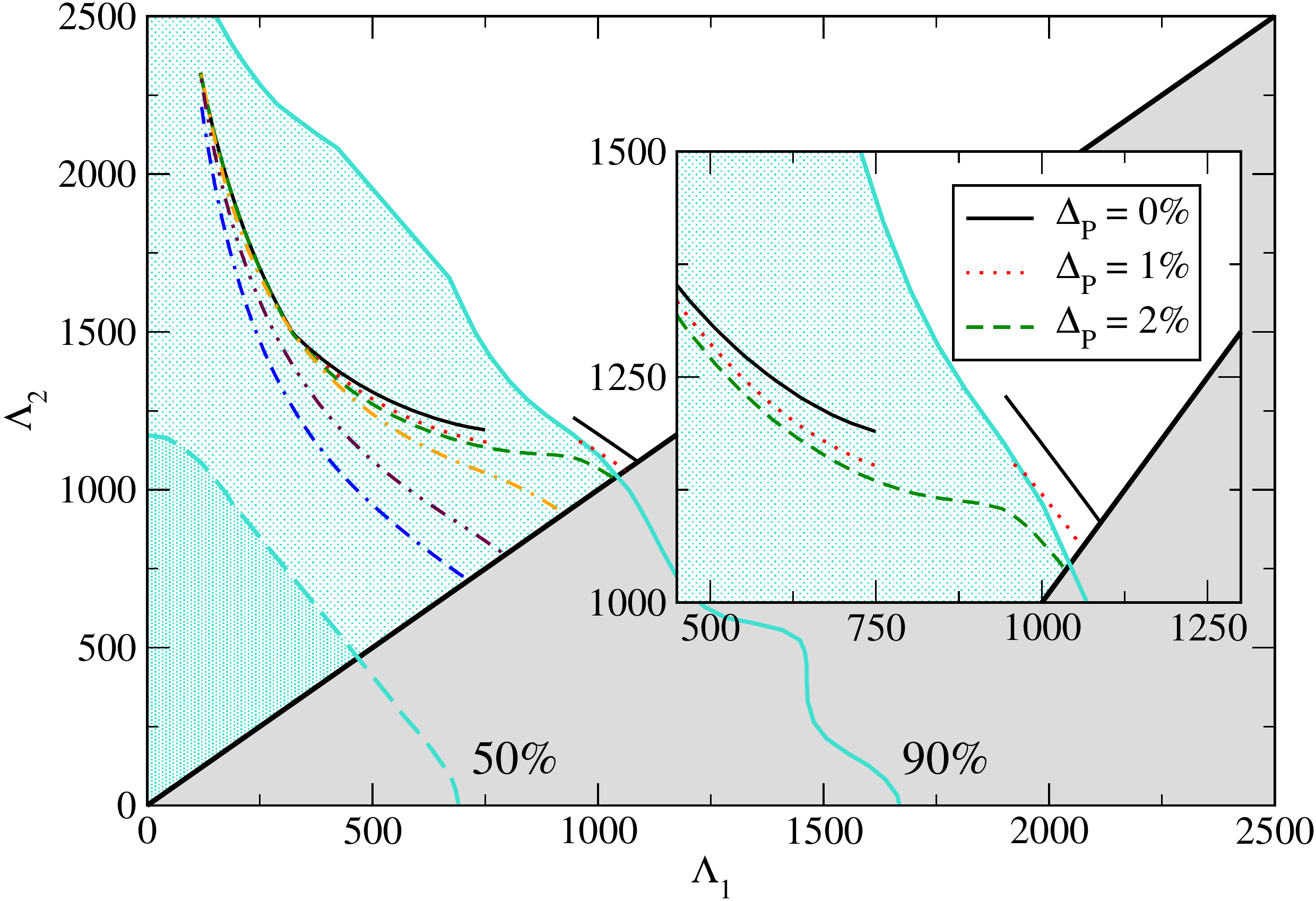}
\end{array}$ 
\par\end{centering}
\caption{ Tidal deformability constraint from GW170817 for equations of state
for high mass twins (left) and low mass twins (right). }
\end{figure*}

Together with the solution of the TOV equations, one can solve for the dimensionless tidal deformability $\Lambda(M)$ for the given EoS.
After that, one can construct the corresponding lines in the $\Lambda_1-\Lambda_2$ diagram of the binary compact star merger GW170817 for which the individual masses $M_1$ and $M_2$ of the two compact stars fulfill the constraint derived from the detected gravitational wave signal of the inspiral phase of the merger \cite{TheLIGOScientific:2017qsa}. 
These lines can be overlaid to the constraint derived from the LVC observation, as shown in Fig.~\ref{L1L2_Pasta}.
As to be expected from Fig.~\ref{fig:MR_ACB4-ACB5_Pasta}, only in the case of ACB5 we can note an effect of the mixed phase construction while the results for ACB4 are inert against changes of the mixed phase parameter, because it influences the mass-radius diagram in a region of masses that is inaccessible to the gravitational wave signal of the inspiral phase and the effects of tidal deformation. 
Moreover, we notice that ACB4 is too stiff an EoS to fulfil the compactness constraint from GW170817.
The EoS ACB5, however, with the early onset of the phase transition, becomes a soft EoS due to the mixed phase effects and  for the largest values of the mixed phase parameter $\Delta_P$ is similar to a soft hadronic EoS despite the fact that the compact stars consist of extended regions of quark matter in pure or mixed phases.

In the following section, we will consider the effects of fast
rotation on the sequences of hybrid star solutions and shall obtain a
qualitative difference in the characteristics of pure phase (hadronic)
and hybrid stars concerning their maximum masses which are relevant
for the discussion of the phenomenology of binary compact star mergers
and their implications for the state of superdense matter.

\subsection{Rotating compact star solutions}
\label{results:rot}
In this subsection we present the numerical solutions for rotating
hybrid star sequences in full GR equations for axial symmetry as
obtained with the RNS code described in subsection \ref{gravity:RNS}
and in the perturbative expansion up to order $\Omega^2$ (slow
rotation approximation) that was explained in subsection
\ref{gravity:mass}.  We relate these solutions to those of the TOV
equations for the static case of spherical symmetry discussed in the
previous subsection \ref{gravity:static}.
\begin{figure*}[!htb]
\begin{centering}
$\begin{array}{cc}
\includegraphics[width=0.5\textwidth]{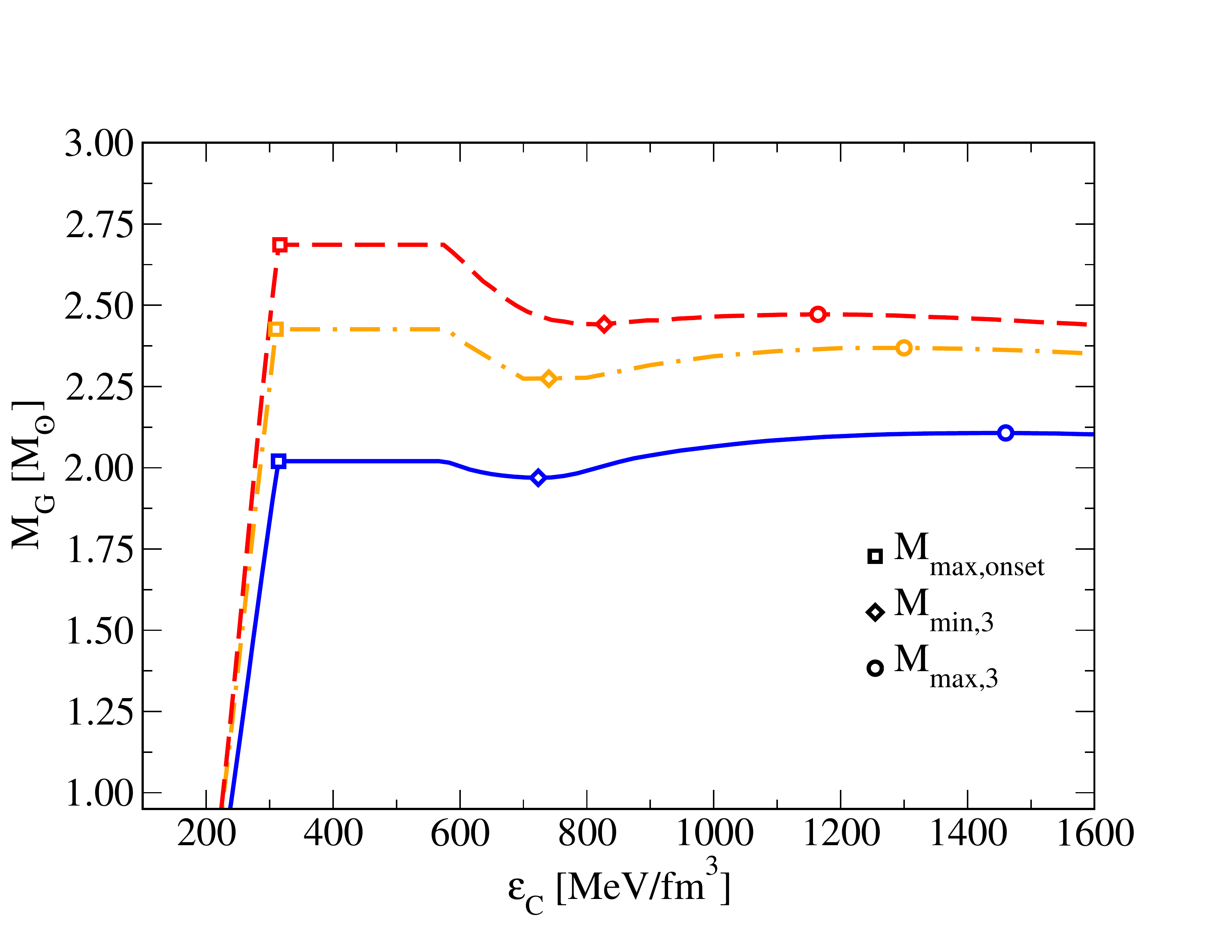} 
& \hspace{0cm}\includegraphics[width=0.5\textwidth]{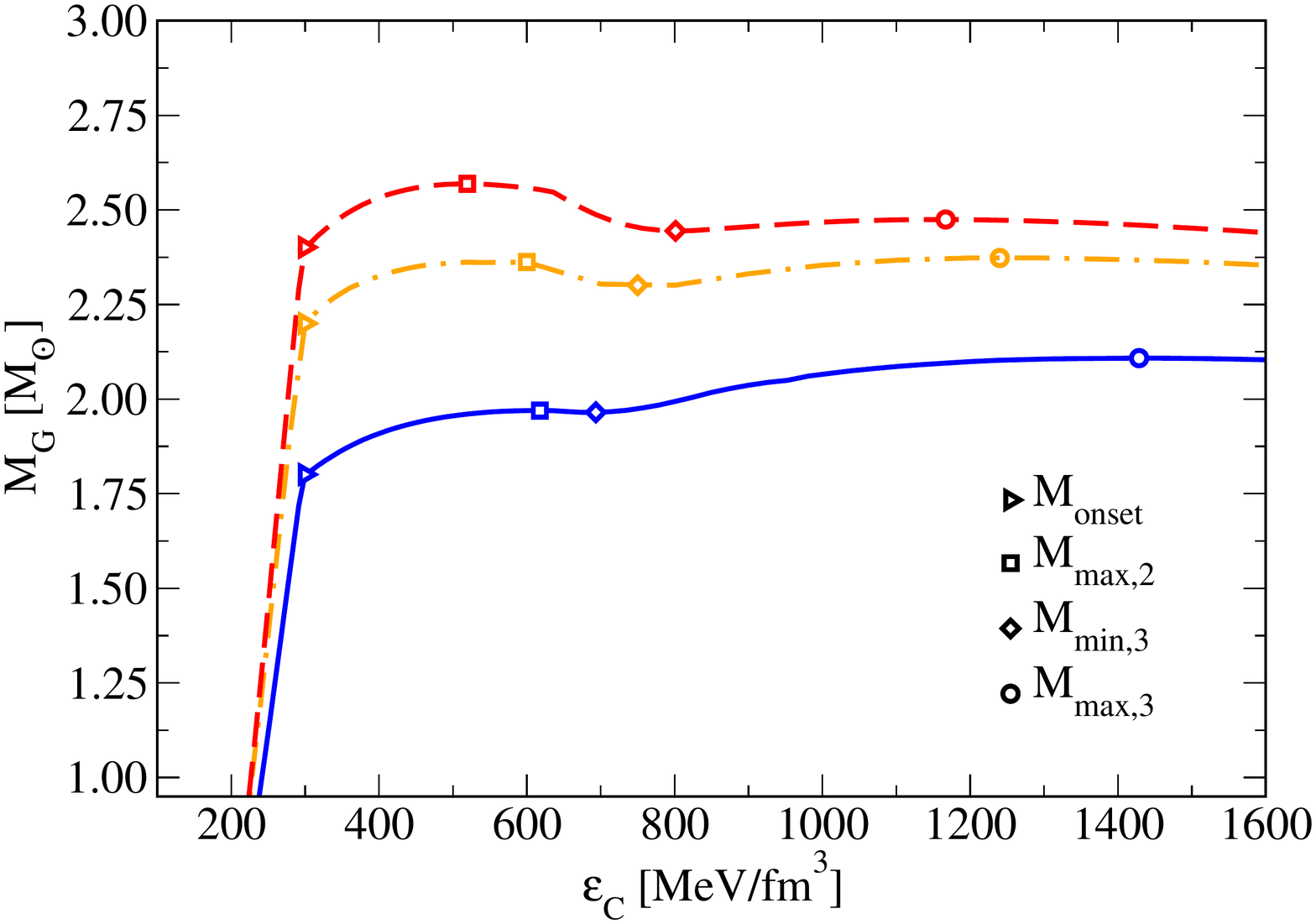}\\[-5mm]
\includegraphics[width=0.5\textwidth]{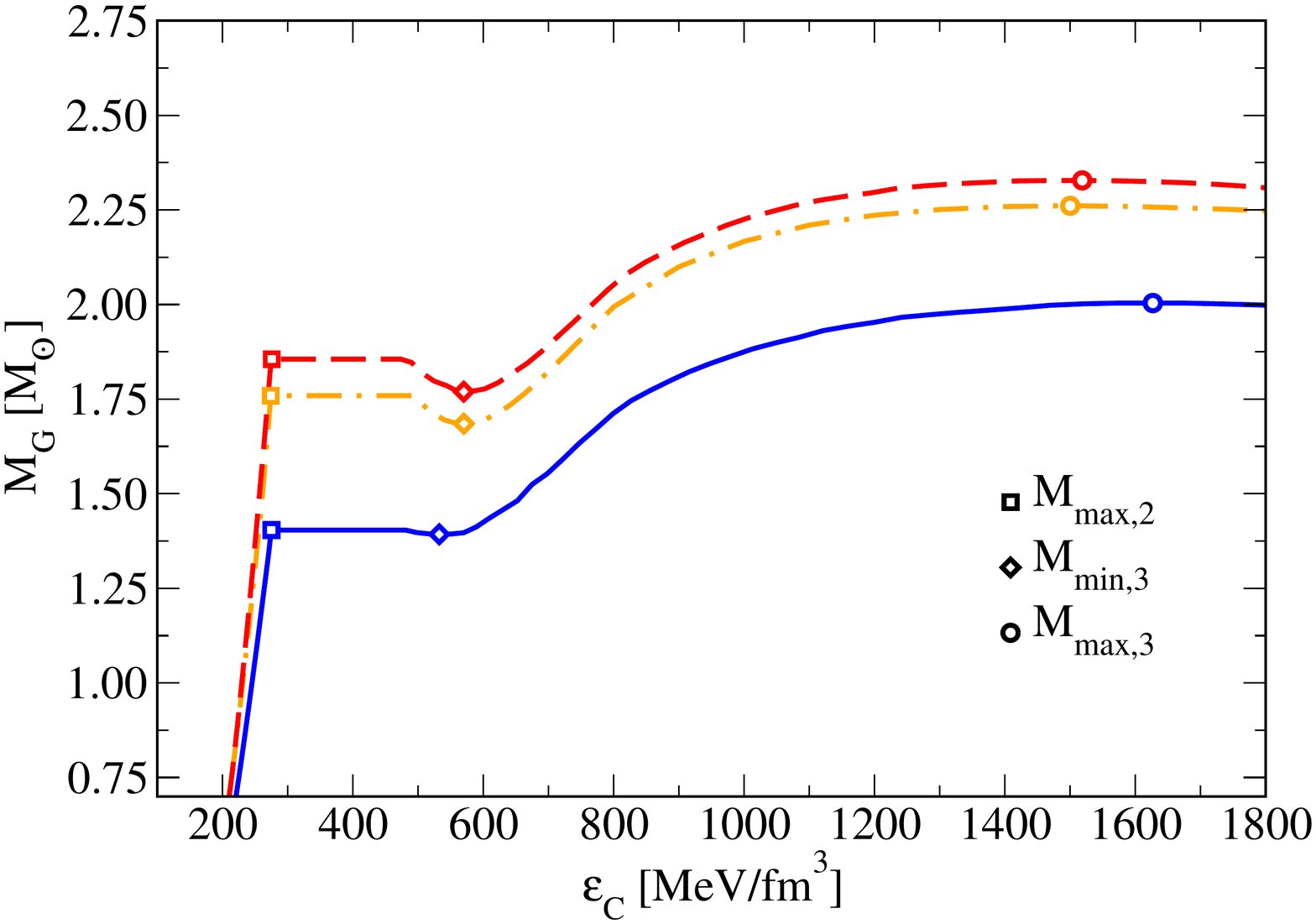} 
& \hspace{0cm}\includegraphics[width=0.5\textwidth]{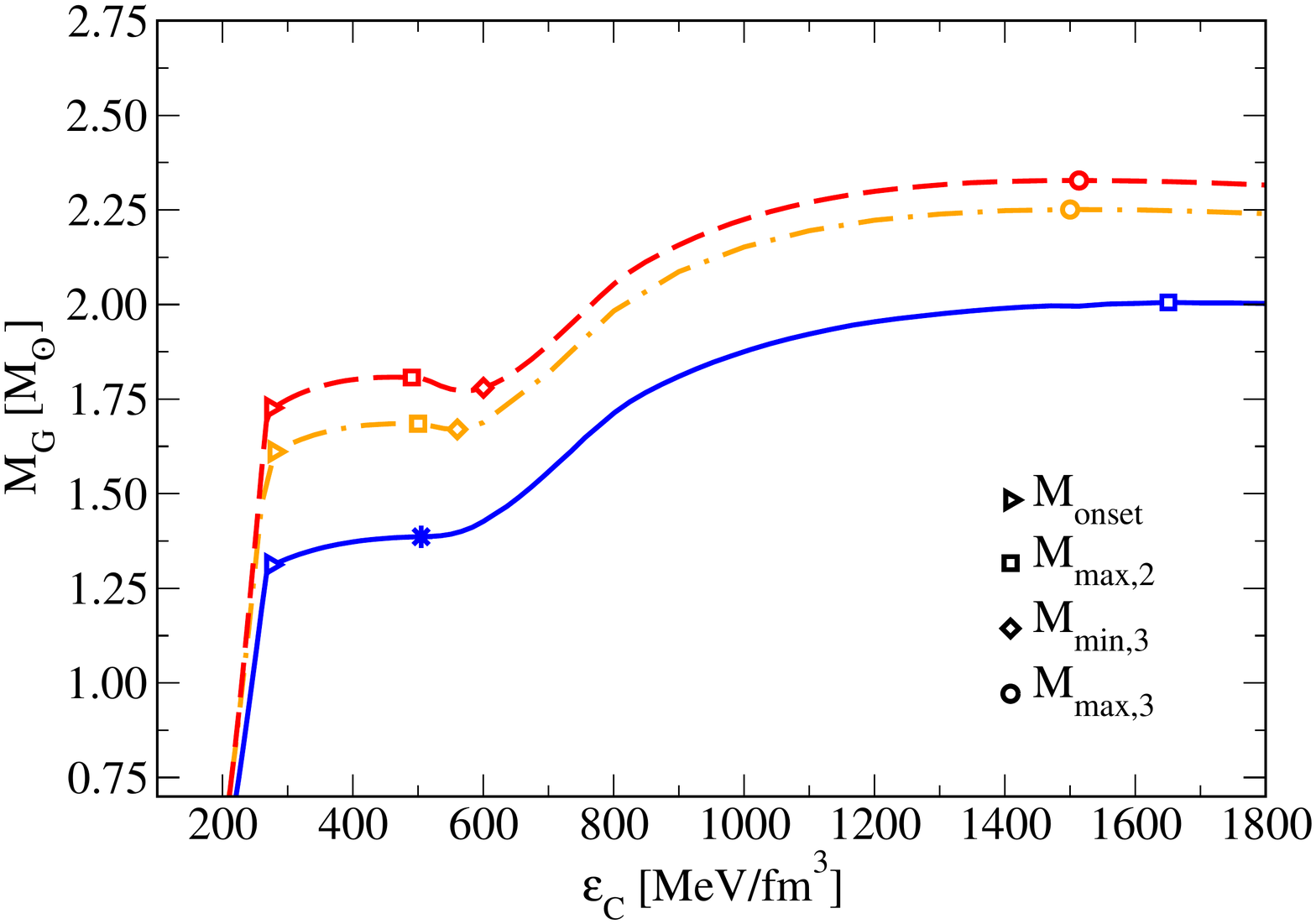}
\end{array}$ 
\par\end{centering}
\caption{\label{fig:ACB4-ACB5:M-e} The Mass--central energy density
  diagram for compact star configurations for the EoS ACB4 (upper
  panels) and ACB5 (lower panels).  The \textit{left} panels are for
  $\Delta_{P}=0$ (Maxwell construction) and the \textit{right} ones
  are for the critical value of the mixed-phase parameter
  ($\Delta_{P}=0.04$ for ACB4 and $\Delta_{P}=0.02$ for ACB5) for
  which the third family vanishes in the static case because it joins
  the second family of neutron stars.  Each panel shows three curves:
  the solution of the TOV equations for the static case (blue solid
  line), the solution of the slow rotation case ($\Omega^2$
  approximation) for rotation at the Kepler frequency $\Omega_K$
  (orange dash-dotted line) and the full solution of the axisymmetric
  Einstein equations with the RNS code for $\Omega=\Omega_K$ (red
  dashed line).  The symbols denote the onset mass for deconfinement
  (triangle right), the maximum mass on the $2^{\rm nd}$ family branch
  (square), the minimum mass (diamond) and the maximum mass (circle)
  on the $3^{\rm rd}$ family branch.  }
\end{figure*}

In Fig.~\ref{fig:ACB4-ACB5:M-e} we show the gravitational mass as a
function of the central energy density for both multipolytrope EoS,
ACB4 (with a deconfinement phase transition at high-mass, upper
panels) and ACB5 (with a transition at the typical compact star mass
of $1.4~M_\odot$, lower panels) for nonrotating (blue solid lines) and
maximally rotating ($\Omega=\Omega_K$) stars in full GR (red dashed
lines) and slow rotation approximation (orange dash-dotted line).  In
the left panels for the Maxwell construction case the jump in the
central energy density by about a factor two at the onset of the
transition is clearly seen and such an amount of latent heat is
sufficient, according to the Seidov criterion Eq.~(\ref{seidov}), to
trigger a gravitational instability which occurs in the region of
densities where
\begin{equation}
\label{instability}
\frac{dM}{d\varepsilon} < 0~.
\end{equation} 
At the onset of this instability the gravitational mass of the star
has reached the maximum attainable on the second family of compact
stars, denoted by $M_{\max,2}$.  For the Maxwell construction case,
this mass is degenerate with that for the onset of the phase
transition, $M_{\rm onset}$.  Due to the absence of a pressure
gradient in the interval of densities corresponding to the mixed
phase, this phase is not realized in compact stars in this case.  The
EoS with mass twin compact star sequences are characterized by the
fact that the instability criterion (\ref{instability}) is fulfilled
in a finite interval of densities which is then followed by another
stable, rising branch of sequences, the so-called third family of
compact stars.  This behavior defines two more characteristic masses:
$M_{\rm min,3}$ at the lower and $M_{\rm max,3}$ at the upper turning
point, see the left panels of Fig.~\ref{fig:ACB4-ACB5:M-e}.

For the mixed phase constructions depicted in the right panels of
Fig.~\ref{fig:ACB4-ACB5:M-e}, the pressures at the onset and the end
of the mixed phase are not identical and thus, due to the
corresponding pressure gradient, a mixed phase can be realized in the
star and the degeneracy between $M_{\rm onset}$ and $M_{\max,2}$ is
lifted.  On the other hand, by our choice of the value of the mixed
phase parameter $\Delta_P$ close to the limiting value for which the
second and the third families of compact stars would get connnected,
the instability vanishes and thus $M_{\rm min,3}$ joins $M_{\max,2}$,
so that for a slightly larger value of $\Delta_P$ both these masses
can no longer be identified since only the second family survives
which for $M>M_{\rm onset}$ consists of hybrid stars.

These four characteristic masses for a given mass twin compact star
EoS are given in Tab.~\ref{tab:TOV} for the static case obtained by
solving the TOV equations. In the last column the absolute maximum of
the mass-radius curve for the given EoS is listed.  In
tables~\ref{tab:Om2} and \ref{tab:RNS} these five characteristic
masses are listed for the sequences of stars rotating at the Kepler
frequency $\Omega_K$ which are obtained from solutions of the
axisymmetric Einstein equations in the $\Omega^2$ approximation and in
full General Relativity, respectively.

\begin{table}[!htb]
\centering 
\tbl{
Five characteristic masses extracted from solutions of the Tolman-Oppenheimer-Volkoff equations (superscript "TOV") for sequences of static configurations with the EoS ACB4 and ACB5 for the Maxwell construction case ($\Delta_P=0$) and for the mixed phase construction with the limiting value of $\Delta_P$ for which the second and the third family branches join. 
$M_{\rm onset}$ is the maximum mass of the purely hadronic second family branch at the onset of deconfinement, $M_{\max_2}$ denotes the maximum mass reached at the end of the mixed phase, $M_{\min_3}$ and $M_{\max_3}$ are the minimum and the  maximum mass on the third family branch of the sequence. The maximum mass of the whole sequence for a given EoS is denoted as $M_{\max}$.}{ 		
		\begin{tabular}{| c | c || c | c | c | c | c |}
			\hline
			&  &  &  &  & \\[-3mm]
			EoS & $\Delta_P$ & $M_{\max,\mathrm{ons}}^{\mathrm{TOV}}$ & $M_{\max,2}^{\mathrm{TOV}}$ & $M_{\min,3}^{TOV}$ & $M_{\max,3}^{TOV}$ & $M_{\max}^{TOV}$ \\[1mm]
			\hline
			\hline
			&  &  &  &  & \\[-3mm]
			\multirow{2}{*}{ACB4} & $0\%$  & 2.020 & 2.020 & 1.969 & 2.107 & 2.107 \\[1mm]
			& $4\%$ & 1.801 & 1.970 & 1.965 & 2.108 & 2.108  \\[1mm]
			\hline
			&  &  &  &  & \\[-3mm]
			\multirow{2}{*}{ACB5} & $0\%$  & 1.404 & 1.404 & 1.393 & 2.004 & 2.004 \\[1mm]
			& $2\%$ & 1.312 & ~\,1.386$^*$ & ~\,1.386$^*$ & ~\,2.006$^*$ & 2.006 \\[1mm]
			\hline
		\end{tabular}
\label{tab:TOV}
}
\end{table}

\begin{table}[!htb]
\centering 
\tbl{
Same as Table \ref{tab:TOV}, but now for solutions of the axisymmetic Einstein equations in the slow rotation approximation, 
the perturbative expansion to order $\Omega^2$ denoted by the corresponding superscript.}{ 		
	\begin{tabular}{| c | c || c | c | c | c | c |}
		\hline
		&  &  &  &  & \\[-3mm]
		EoS & $\Delta_P$ & $M_{\max,\mathrm{ons}}^{\Omega^2}$ & $M_{\max,2}^{\Omega^2}$ & $M_{\min,3}^{\Omega^2}$ & $M_{\max,3}^{\Omega^2}$ & $M_{\max}^{\Omega^2}$ \\[1mm]
		\hline
		\hline
		&  &  &  &  & \\[-3mm]
		\multirow{2}{*}{ACB4} & $0\%$  & 2.426 & 2.426 & 2.274 & 2.369 & 2.426 \\[1mm]
		& $4\%$ & 2.200 & 2.362 & 2.301 & 2.373 & 2.373  \\[1mm]
		\hline
		&  &  &  &  & \\[-3mm]
		\multirow{2}{*}{ACB5} & $0\%$  & 1.759 & 1.759 & 1.685 & 2.261 & 2.261 \\[1mm]
		& $2\%$ & 1.611 & 1.685 & 1.670 & 2.251 & 2.251 \\[1mm]
		\hline
	\end{tabular}
\label{tab:Om2}
}
\end{table}

\begin{table}[!htb]
\centering 
\tbl{
Same as Table \ref{tab:TOV}, but now for solutions of the the full system of Einstein equations for uniform rotation in axial symmetry 
\cite{Cook:1993qj} using the RNS code. The corresponding results are denoted by the superscript "rot".}
{ 		
	\begin{tabular}{| c | c || c | c | c | c | c |}
		\hline
		&  &  &  &  & \\[-3mm]
		EoS & $\Delta_P$ & $M_{\mathrm{onset}}^{\mathrm{rot}}$ & $M_{\max,2}^{\mathrm{rot}}$ & $M_{\min,3}^{rot}$ & $M_{\max,3}^{rot}$ & $M_{\max}^{rot}$ \\[1mm]
		\hline
		\hline
		&  &  &  &  & \\[-3mm]
		\multirow{2}{*}{ACB4} & $0\%$  & 2.686 & 2.686 & 2.442 & 2.472 & 2.686 \\[1mm]
		& $4\%$ & 2.401 & 2.569 & 2.445 & 2.475 & 2.569  \\[1mm]
		\hline
		&  &  &  &  & \\[-3mm]
		\multirow{2}{*}{ACB5} & $0\%$  & 1.855 & 1.855 & 1.770 & 2.328 & 2.328 \\[1mm]
		& $2\%$ & 1.727 & 1.807 & 1.780 & 2.328 & 2.328 \\[1mm]
		\hline
	\end{tabular}
\label{tab:RNS}
}
\end{table}

Inspecting the rotating star sequences for the close-to-critical mixed phase parameter $\Delta_P$, we observe that due to the rotation the star branches with mixed phase and pure quark matter core can get disconnected so that the phenomenon of a third family reappears.
Vice-versa, upon spin-down from a supramassive star configuration at maximal rotation frequency (created, e.g., in a binary neutron star merger) which is stable on the hadronic or mixed phase branch may end up either as a black hole or on the hybrid star branch for such mixed-phase EoS.   
The scenario of a delayed collapse to a black hole is of special importance for interpreting GW170817 and will therefore be discussed below in further detail. 
In this context appears the question whether between the above-introduced characteristic masses at maximal and at zero rotation frequency hold EoS-independent, so-called {\it universal relations} that have been investigated for the maximum mass of hadronic EoS \cite{Bozzola:2017qbu,Rezzolla:2017aly} and recently also for hybrid EoS including the Maxwell construction cases of ACB4 and ACB5 
\cite{Bozzola:2019tit}. We shall come back to this issue below.

\begin{figure*}[!hbt]
\begin{centering}
$\begin{array}{cc}
\includegraphics[width=0.5\textwidth]{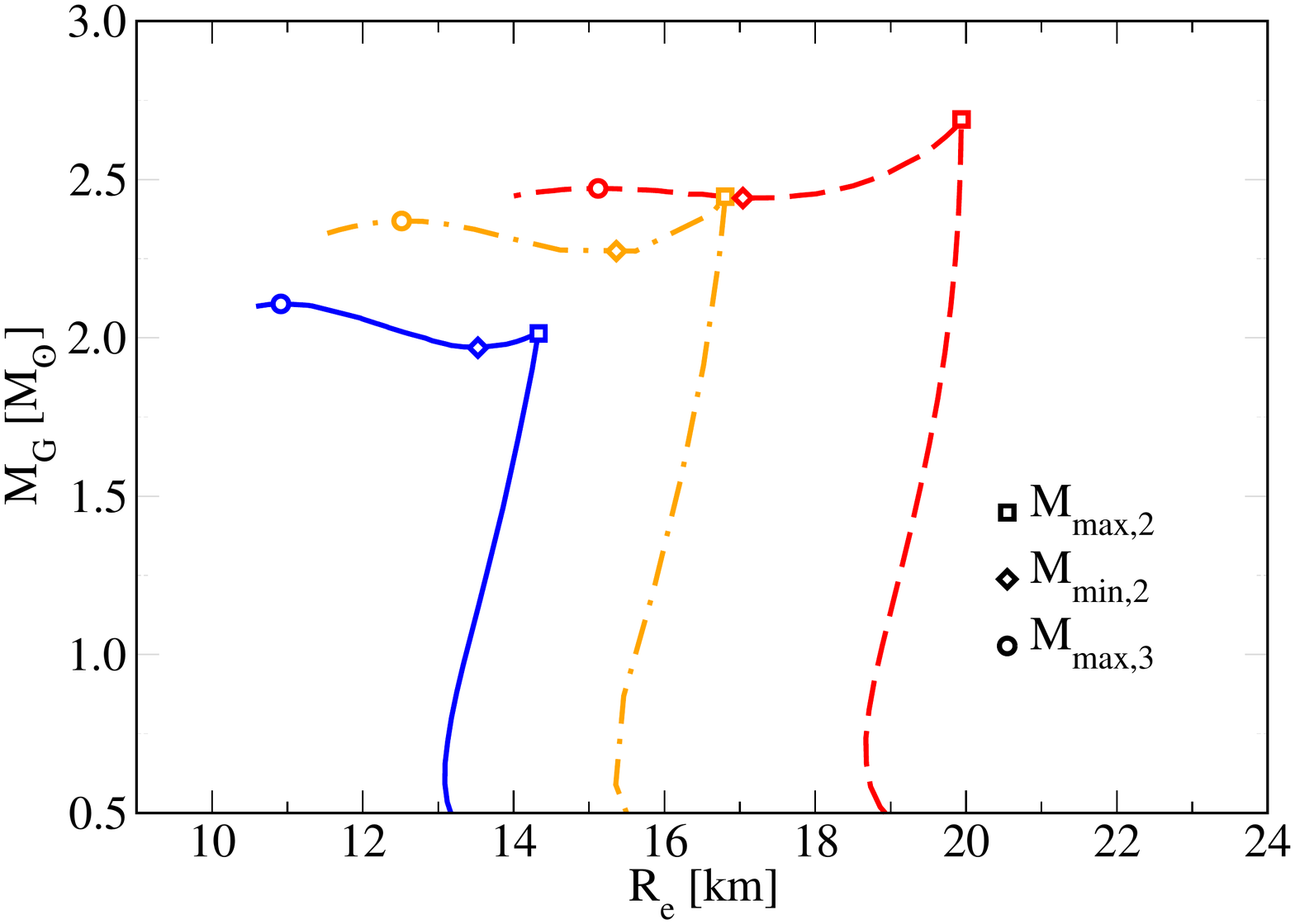} 
& \hspace{0cm}\includegraphics[width=0.5\textwidth]{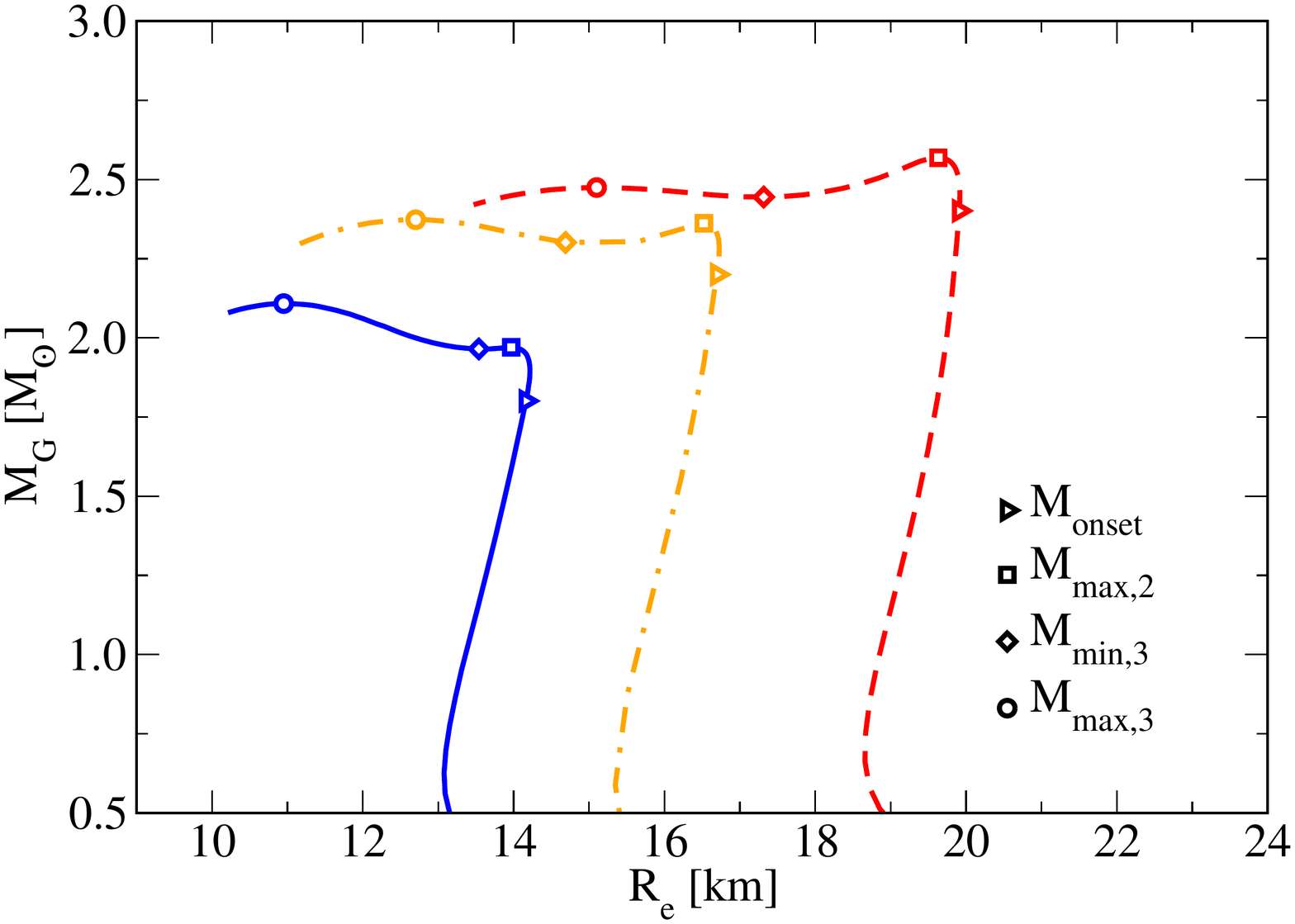}\\[-5mm]
\includegraphics[width=0.5\textwidth]{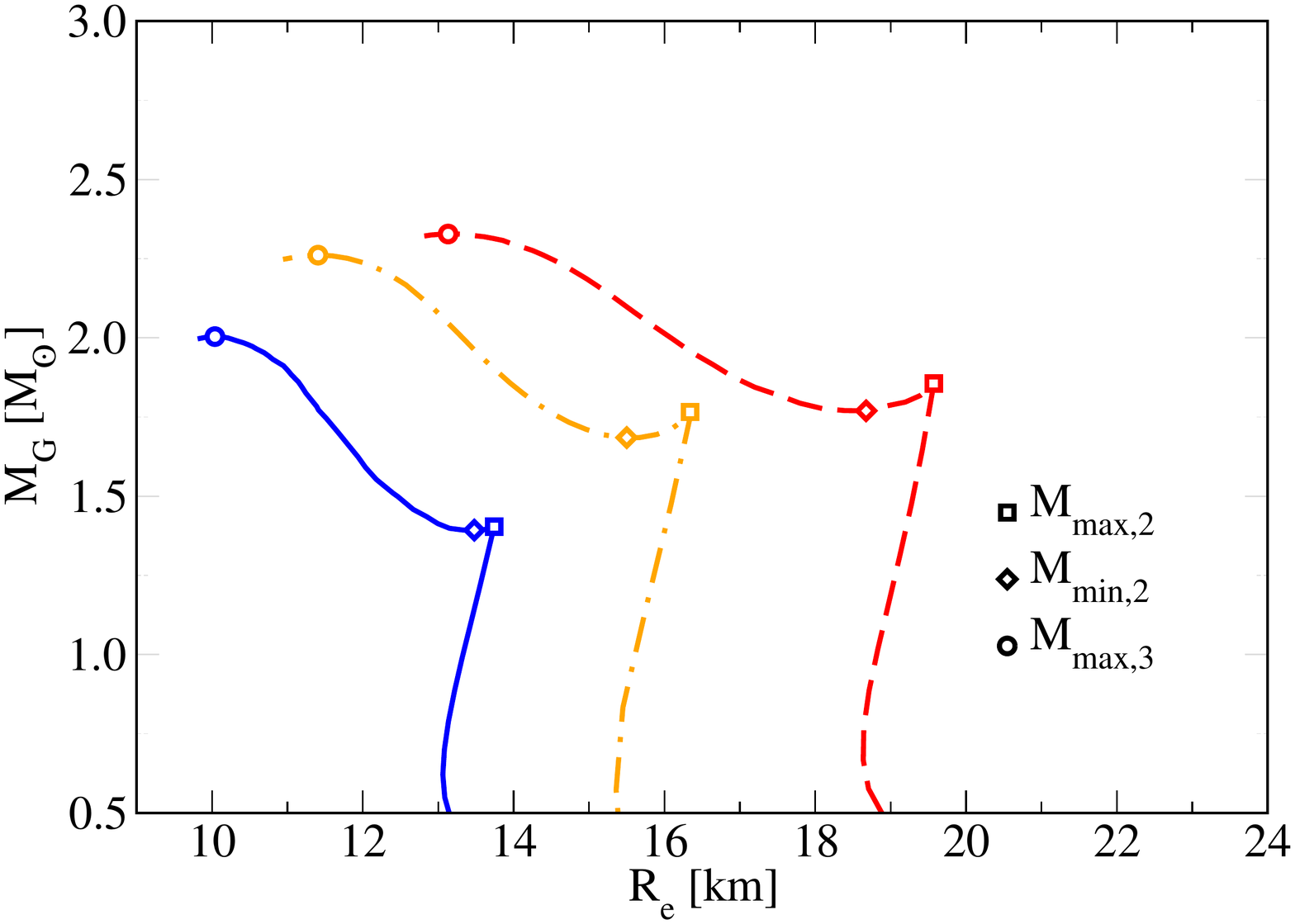} 
& \hspace{0cm}\includegraphics[width=0.5\textwidth]{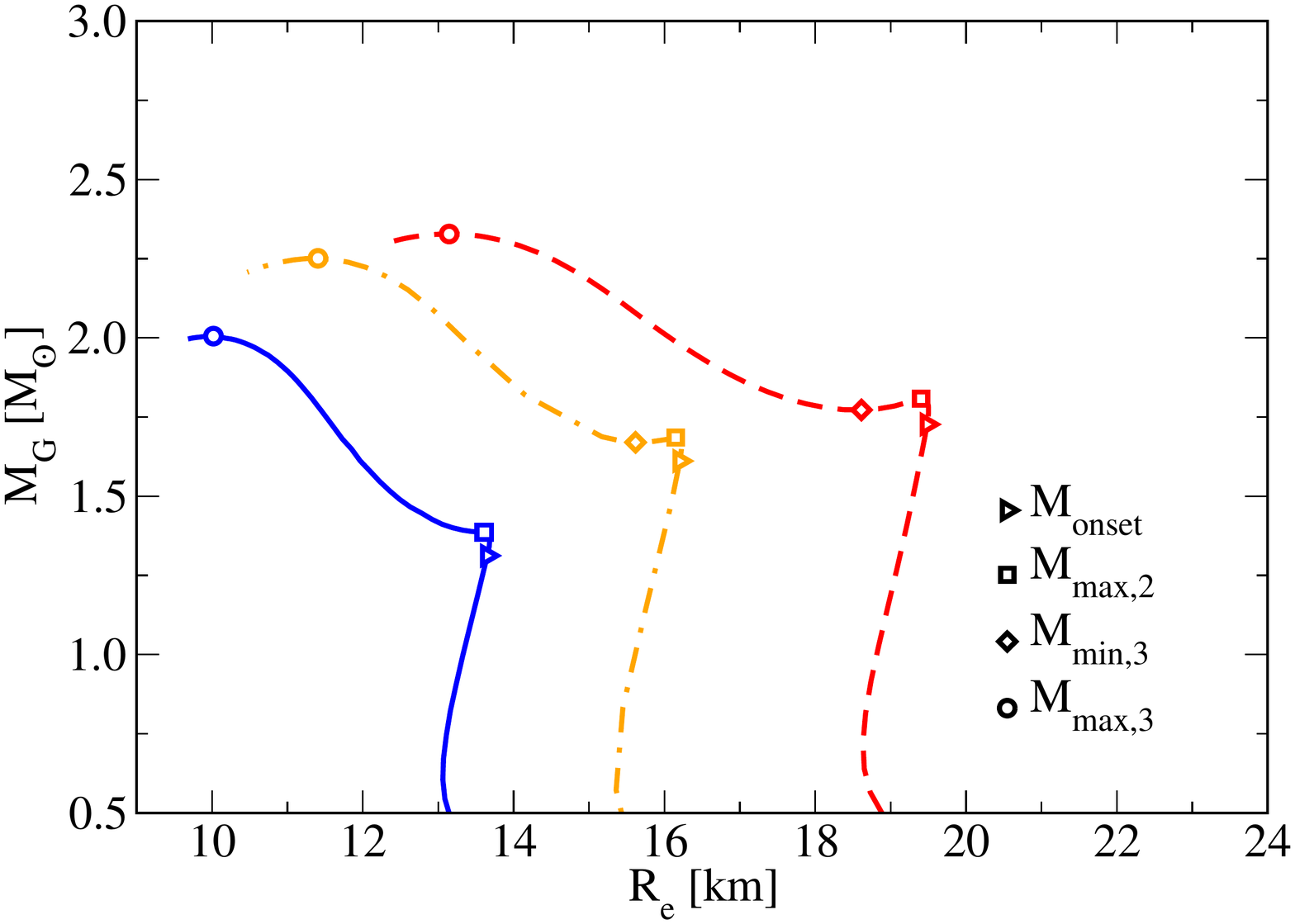}
\end{array}$ 
\par\end{centering}
\caption{\label{fig:ACB4-ACB5:MR} The Mass--Radius diagram for compact star configurations for 
the EoS ACB4 (upper panels) and ACB5 (lower panels).
The \textit{left} panels are for $\Delta_{P}=0$ (Maxwell construction) and
the \textit{right} ones are for the critical value of the mixed-phase parameter ($\Delta_{P}=0.04$ for ACB4 and 
$\Delta_{P}=0.02$ for ACB5) for which the third family 
vanishes in the static case because it joins the second family of neutron stars. 
Each panel shows three curves: the solution of the TOV equations for the static case (blue solid line), the solution of the slow 
rotation case ($\Omega^2$ approximation) for rotation at the Kepler frequency $\Omega_K$ (orange dash-dotted line) and the 
full solution of the axisymmetric Einstein equations with the RNS code for $\Omega=\Omega_K$ (red dashed line).
The symbols denote the onset mass for deconfinement (triangle right), the maximum mass on the $2^{\rm nd}$ family branch 
(square), the minimum mass (diamond) and the maximum mass (circle) on the $3^{\rm rd}$ family branch.  }
\end{figure*}

Here we like to remark that the maximum masses on the second and third family branches correspond to stars with very different central (energy) densities. This may be the clue to understanding the fact that the increase in mass for stars on the more compact third family branch is smaller than for stars on the secnd family one because of their smaller radii and thus smaller moment of inertia (\ref{eq:DI}) and rotational energy, see the gravitational mass vs. equatorial radius in Fig.~\ref{fig:ACB4-ACB5:MR}.
For a more quantitative discussion of the effects of rotation on the masses of the sequences, we extract from the 
tables~\ref{tab:TOV}-\ref{tab:RNS} the ratios of the characteristic masses on the rotating sequences to those on the static ones in 
table~\ref{tab:Om2-TOV} and table~\ref{tab:RNS-TOV} for the $\Omega^2$ approximation and the full GR solution, respectively.
For completeness, we give the ratio of the characteristic masses between the two rotation solutions in  table~\ref{tab:RNS-Om2}.


\begin{table}[!b]
\centering 
\tbl{
Ratios formed by the five charactaristic masses at maximal rotation frequency $\Omega_K$ in the slow rotation approximation 
(superscript "$\Omega^2$") relative to the static case (superscript "TOV") calculated with the EoS ACB4 and ACB5 for the Maxwell construction case ($\Delta_P=0$) and for the mixed phase construction with the limiting value of $\Delta_P$.
}{ 		
	\begin{tabular}{| c | c || c | c | c | c | c |}
		\hline
		&  &  &  &  & & \\[-3mm]
		EoS & $\Delta_P$ & $\displaystyle \frac{M_{\mathrm{ons}}^{\Omega^2}}{M_{\mathrm{ons}}^{\mathrm{TOV}}}$ & $\displaystyle \frac{M_{\max,2}^{\Omega^2}}{M_{\max,2}^{\mathrm{TOV}}}$ & $\displaystyle \frac{M_{\min,3}^{\Omega^2}}{M_{\min,3}^{\mathrm{TOV}}}$ & $\displaystyle \frac{M_{\max,3}^{\Omega^2}}{M_{\max,3}^{\mathrm{TOV}}}$ & $\displaystyle \frac{M_{\max}^{\Omega^2}}{M_{\max}^{\mathrm{TOV}}}$ \\[3mm]
		\hline
		\hline
		&  &  &  &  & \\[-3mm]
		\multirow{2}{*}{ACB4} & $0\%$  & 1.201 & 1.201 & 1.155 & 1.124 & 1.151 \\[1mm]
		& $4\%$ & 1.222 & 1.199 & 1.171 & 1.126 & 1.126 \\[1mm]
		\hline
		&  &  &  &  & & \\[-3mm]
		\multirow{2}{*}{ACB5} & $0\%$  &  1.253 & 1.253 & 1.210 & 1.128 & 1.128 \\[1mm]
		& $2\%$ & 1.228 & 1.216$^*$ & 1.205$^*$ & 1.122$^*$ & 1.122 \\[1mm]
		\hline
	\end{tabular}
\label{tab:Om2-TOV}
}
\end{table}

\begin{table}[!b]
\centering 
\tbl{
Same as Table \ref{tab:Om2-TOV}, but now for the full solutions of the axisymmetic Einstein equations with the RNS code 
relative to the solutions of the TOV equations for the static case.}{ 		
	\begin{tabular}{| c | c || c | c | c | c | c |}
		\hline
		&  &  &  &  & & \\[-3mm]
		EoS & $\Delta_P$ & $\displaystyle \frac{M_{\mathrm{ons}}^{\mathrm{rot}}}{M_{\mathrm{ons}}^{\mathrm{TOV}}}$ & $\displaystyle \frac{M_{\max,2}^{\mathrm{rot}}}{M_{\max,2}^{\mathrm{TOV}}}$ & $\displaystyle \frac{M_{\min,3}^{\mathrm{rot}}}{M_{\min,3}^{\mathrm{TOV}}}$ & $\displaystyle \frac{M_{\max,3}^{\mathrm{rot}}}{M_{\max,3}^{\mathrm{TOV}}}$ & $\displaystyle \frac{M_{\max}^{\mathrm{rot}}}{M_{\max}^{\mathrm{TOV}}}$ \\[3mm]
		\hline
		\hline
		&  &  &  &  & \\[-3mm]
		\multirow{2}{*}{ACB4} & $0\%$  & 1.330 & 1.330 & 1.240 & 1.173 & 1.275 \\[1mm]
		& $4\%$ & 1.333 & 1.304 & 1.244 & 1.174 & 1.174 \\[1mm]
		\hline
		&  &  &  &  & & \\[-3mm]
		\multirow{2}{*}{ACB5} & $0\%$  &  1.321 & 1.321 & 1.270 & 1.162 & 1.162 \\[1mm]
		& $2\%$ & 1.316 & 1.304$^*$ & 1.284$^*$ & 1.161$^*$ & 1.161 \\[1mm]
		\hline
	\end{tabular}
\label{tab:RNS-TOV}
}
\end{table}

\begin{table}[!b]
\centering 
\tbl{
Same as Table \ref{tab:Om2-TOV}, but now the full solutions of the axisymmetic Einstein equations (superscript "RNS") are related to those in the slow rotation approximation (superscript $\Omega^2$).}{ 		
	\begin{tabular}{| c | c || c | c | c | c | c |}
		\hline
		&  &  &  &  & & \\[-3mm]
		EoS & $\Delta_P$ & $\displaystyle \frac{M_{\mathrm{ons}}^{\mathrm{rot}}}{M_{\mathrm{ons}}^{\Omega^2}}$ & $\displaystyle \frac{M_{\max,2}^{\mathrm{rot}}}{M_{\max,2}^{\Omega^2}}$ & $\displaystyle \frac{M_{\min,3}^{\mathrm{rot}}}{M_{\min,3}^{\Omega^2}}$ & $\displaystyle \frac{M_{\max,3}^{\mathrm{rot}}}{M_{\max,3}^{\Omega^2}}$ & $\displaystyle \frac{M_{\max}^{\mathrm{rot}}}{M_{\max}^{\Omega^2}}$ \\[3mm]
		\hline
		\hline
		&  &  &  &  & \\[-3mm]
		\multirow{2}{*}{ACB4} & $0\%$  & 1.107 & 1.107 & 1.074 & 1.043 & 1.107 \\[1mm]
		& $4\%$ & 1.091 & 1.087 & 1.063 & 1.043 & 1.082 \\[1mm]
		\hline
		&  &  &  &  & & \\[-3mm]
		\multirow{2}{*}{ACB5} & $0\%$  &  1.054 & 1.054 & 1.050 & 1.030 & 1.030 \\[1mm]
		& $2\%$ & 1.072 & 1.072 & 1.065 & 1.034 & 1.034 \\[1mm]
		\hline
	\end{tabular}
\label{tab:RNS-Om2}
}
\end{table}

\section{Implications for the phenomenology of compact stars 
\label{Implic}}

We investigate the consequences of a strong phase transition in the
EoS for dense compact star matter for sequences of configurations in
the mass-radius as well as mass-central (energy) density plane, with
and without rotation.  While for isolated pulsars even the highest
known spin frequencies are well below the Kepler frequency so that no
strong modification of the TOV solution occurs, in the era of
multi-messenger astronomy, with compact star mergers being accessible
to detection by their GW signal from the inspiral phase and soon also
from the postmerger state, the "\"ubermassive" \cite{Espino:2019ebx}
as well as supramassive star solutions play a role in the
interpretation of the observations. While the former are solutions for
differentially rotating configurations at the mass shedding limit, the
latter are uniformly rotating objects with a frequency close to the
Kepler one \cite{Shibata:2019ctb}.  Recently, for this case EoS
independent, so-called "universal" relationships have been derived
which relate the maximum mass of the supramassive star sequence to the
static one from the solution of the TOV equation.  Such relationships,
once confirmed, are particularly useful in order to make general
predictions or draw conclusions from merger phenomenology for
constraints limiting the EoS properties.  In this context we would
like to mention the upper limit on the maximum mass of neutron stars
that has been extracted from the GW signal and phenomenology of
GW170817 \cite{Shibata:2017xdx,Margalit:2017dij,Rezzolla:2017aly}.

\begin{figure*}[!hbt]
\begin{centering}
\includegraphics[width=0.7\textwidth]{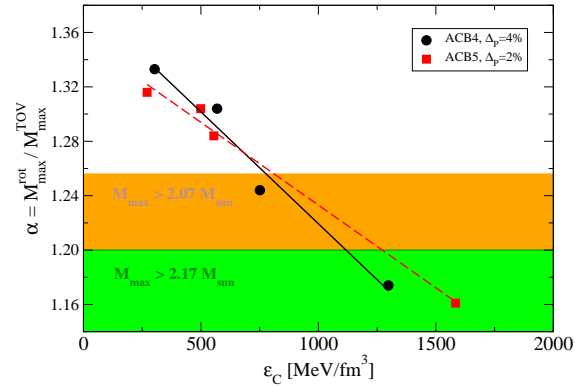} 
\par\end{centering}
\caption{\label{fig:ratio} 
Ratio of the mass for maximally rotating stars to that of a static star as a function of the central energy density.
For a comparison the value 1.20 would be the maximum value compatible with a lower limit on the maximum mass 
of (nonrotating) pulsars of $2.17~M_\odot$ \cite{Cromartie:2019kug}. }
\end{figure*}

It has been known since long \cite{Haensel:2007yy} that there is a
relationship between the maximum mass of uniformly rotating cold
neutron stars at the maximum frequency and the maximum mass of the TOV
equation solution for static stars as
\begin{equation}
\label{Mmax}
M_{\rm max}(\Omega_K)=\alpha M_{\rm max}^{\rm TOV}~,
\end{equation}
where recently in Ref.~\cite{Breu:2016ufb} the universality of this
relationship was confirmed for a very large set of hadronic EoS
(without a deconfinement phase transition) with the coefficient
$\alpha=1.20$.  The hypothesis that the relation (\ref{Mmax}) can be
extended to include hybrid stars with a strong phase transition and
even with third family sequences has recently been investigated in
Ref.~\cite{Bozzola:2019tit} and following the argumentation of
\cite{Ruiz:2017due,Most:2018hfd,Rezzolla:2017aly} it leads to a
limitation for the maximum mass of static neutron stars as
\begin{equation}
\label{Mmax-limit}
2.07~M_\odot \simeq M_{\rm max}^{\rm TOV}\simeq 2.23~M_\odot~.
\end{equation}
The value of $M_{\rm max}=2.591~M_\odot$ was extracted for the 
core mass of the compact star merger GW170817 in Ref.~\cite{Rezzolla:2017aly}.
We confirm the finding of \cite{Bozzola:2019tit} that the coefficient
spans a range of values for which we find $1.16 < \alpha < 1.33$, see
the table \ref{tab:RNS-TOV}.  
The lower limit in (\ref{Mmax-limit}) comes from the new high-mass pulsar PSR J0740+6620 for which
a mass $2.17^{+0.11}_{-0.10}~M_\odot$ has been determined by~\citet{Cromartie:2019kug} 
by measuring the Shapiro delay, and the upper limit in our case is $2.23~M_\odot$ 
for the lowest value of $\alpha=1.16$.
In order to not come in conflict with the pulsar mass measurement of
\citep{Cromartie:2019kug}, there is an upper limit for the admissible 
value of $\alpha=1.25$, corresponding to the lower limit at the $1\sigma$ level 
of the PSR J0740+6620 mass, $M_{\rm max}^{\rm TOV}=2.07~M_\odot$.
Taking the central value,  $M_{\rm max}^{\rm TOV}=2.17~M_\odot$, would correspond to 
$\alpha=1.20$, see Fig.~\ref{fig:ratio}.
This figure illustrates one of the main findings of this contribution, the dependence
of the coefficient $\alpha$ in Eq.~(\ref{Mmax}) on the central (energy) density of the
stellar configuration that can be fitted by a linear regression to the values we determined
at the positions corresponding to the characteristic masses to be
\begin{equation}
\label{alpha}
\alpha=a - b \varepsilon_c~,
\end{equation}
where $a_{ACB4-4}=1.38\pm 0.07$, $b_{ACB4-4}=0.12\pm 0.01$ fm$^3$/GeV and
$a_{ACB5-2}=1.37\pm 0.07$, $b_{ACB5-2}=0.16\pm 0.02$ fm$^3$/GeV.

We may conclude that only those states of matter are allowed for the inner core of 
a compact star at maximum mass of $2.07~M_\odot$ ($2.17~M_\odot$) which belong to 
a high-density region with $\varepsilon\ge 0.78$ GeV/fm$^3$ ($\varepsilon\ge 1.12$ GeV/fm$^3$).

Anyway, the main effect of the strong phase transition is a higher
compactness of the high mass stars than in the purely hadronic case
which moreover goes along with a smaller mass increase due to maximal
rotation than in the purely hadronic case which entails the increase
of the upper limit for the maximum mass relative to the purely
hadronic case discussed in
\cite{Shibata:2017xdx,Margalit:2017dij,Rezzolla:2017aly}. Thus the
high-density phase transition removes a certain tension from the
discussion of the upper limit for the maximum mass and could be used
as an argument in favour of the suggestion that a strong phase
transition actually takes place in compact stars!

\section{Summary and Conclusions}
\label{discussion}

Stimulated by the unprecedented progress in observational astronomy,
compact stars have become superb astrophysical laboratories for a
broad range of physical studies. 
{
This is particularly the case for neutron stars,
since their observables carry information about the fundamental
building blocks of matter and even of the fabric of space
itself. Against this background we did present in this book chapter a
systematic investigation of the properties of compact stellar mass
twins (i.e., the so-called third family of compact stars), which,
according to theory, may exist in the mass-radius region between
neutron stars and stellar-mass black holes. Particular emphasis is
given to modeling the rotational properties of compact mass twins for
multi-polytrope models for the equation of state of ultra-dense
stellar matter that fulfill the constraint established for the maximum
mass of a neutron star.  The main results of our investigation can be
summarized as follows:

1) The existence of mass twins invariably signals the existence of a
strong phase transition in ultra-dense matter.  The extreme softening
of the equation of state caused by the strong phase transition
increases the gravitational field so much that the star becomes
gravitationally unstable over a certain range of densities, where just
a certain fraction of the star's core is in the new high-density
phase.  Eventually the stars becomes stable again when about half of
the matter in its core is in the new phase of matter. The new stars
have gravitational masses that are less than the maximal mass of the
hadronic stellar sequence.

The observation of mass twins would indicate a strong phase transition
so that from the existence of a strong first-order phase transition
in one corner of the QCD phase diagram and a crossover behavior in
another, one could conclude that at least one critical endpoint (CEP) must
exist.  This would be very reassuring for large scale experimental
heavy-ion collision programs set up for the search for the CEP.

2) The mixed phase construction which mimics the pasta phase is in
accordance with a full pasta calculation. The result is a "smearing"
of the phase transition over a certain pressure region, which is
similar to the Gibbs construction in matter with more than one conserved charge
and where charge conservation need not be fulfilled locally
but rather globally \cite{Glendenning:1992vb}.  This construction makes
not only the approach to the strong phase transition more realistic,
but  has also great advantages for the numerical realization of
phenomenological scenarios of the phase transition in rapidly rotating
stars studied in numerical relativity.

3) Lastly, we address the conjecture of an upper limit on the maximum
masses of nonrotating compact stars from the phenomenology of GW170817
and its associated kilonova event.  The conjecture is based on a
quasi-universal relation between the maximum masses of uniformly
rotating stars at the maximum frequency and that of static TOV
solutions for the same EoS, which was demonstrated to hold for neutron
star EoS without a phase transition \cite{Breu:2016ufb}.
The stellar mass at which the high-density phase transition (such as
deconfinement of quarks) sets in is currently unknown. But if the
transition would occur below the maximum mass of the TOV solution, the
quasi-universal relation will have to be revisited and the conclusions
for the upper limit on the maximum mass be revised.  We have found a
quantitative criterion for the minimal central energy density in the
maximum-mass configuration of a compact star that would correspond to
the core of GW170817 after dynamical mass ejection \cite{Rezzolla:2017aly}.  
Thus the EoS at high densities must be effectively soft, either in the form
of a relatively soft hadronic EoS or as a hybrid EoS with a phase
transition, since a too stiff a hadronic EoS would lead to heavy hadronic stars
too dilute to fulfill the constraint derived by us.

With these prospects for strong gravity and strong phase transitions
in compact stars, we are looking forward to the next series of exciting 
discoveries in the just opened era of multimessenger astronomy.

\subsection*{Acknowledgements}
\label{thanks}

We would like to thank Andreas Bauswein for helpful discussions and 
Vasilis Paschalidis for pointing out caveats in the discussion of the maximum mass 
constraint from GW170817.
We are grateful to Cesar Zen Vasconcellos for the invitation to contribute to this book
and for his steady encouragement to complete the writeup despite interfering obligations.
The work of D.B., A.A., H.G. has been supported by the Russian Science Foundation 
under grant 17-12-01427; F.W. acknowledges support by the U.S. National Science 
Foundation under Grant PHY-1714068. 
The authors are grateful to the European COST Actions CA15213 "THOR" and 
CA16214 "PHAROS" for supporting their networking activities.

\end{document}